\documentclass[aps,pre,reprint,superscriptaddress,footinbib,floatfix]{revtex4-2}

\usepackage{graphicx}
\usepackage{bm}
\usepackage{amssymb}
\usepackage{amsbsy}
\usepackage{amsmath}
\usepackage{mathtools}
\usepackage{xcolor}
\usepackage{stmaryrd}
\usepackage{mathrsfs}
\usepackage[mathscr]{euscript}
\usepackage{tikz}
\usepackage{float}
\usetikzlibrary{shapes}
\usepackage{cancel}
\usepackage[normalem]{ulem}

\begin{document}


\title{Dynamics of a spherical self-propelled tracer in a polymeric medium: interplay of self-propulsion, stickiness, and crowding}


\author{Ramanand Singh Yadav}
\affiliation{Department of Chemistry, Indian Institute of Technology Bombay, Mumbai, Maharashtra -  400076, India}
\author{Chintu Das}
\affiliation{Department of Chemistry, Indian Institute of Technology Bombay, Mumbai, Maharashtra -  400076, India}
\author{Rajarshi Chakrabarti}
\email{rajarshi@chem.iitb.ac.in}
\affiliation{Department of Chemistry, Indian Institute of Technology Bombay, Mumbai, Maharashtra -  400076, India}



\begin{abstract}
\noindent We employ computer simulations to study the dynamics of a self-propelled spherical tracer particle in a viscoelastic medium, made of a long polymer chain. Here, the interplay between viscoelasticity, stickiness, and activity (self-propulsion) brings additional complexity to the tracer dynamics. Our simulation shows that on increasing the stickiness of the tracer particle to the polymer beads, the dynamics of the tracer particle slows down as it gets stuck to the polymer chain, and moves along with it. But with increasing the self-propulsion velocity, the dynamics gets enhanced. In case of increasing stickiness as well as activity, the non-Gaussian parameter (NGP) exhibits non-monotonic behavior, which also shows up in the re-scaled self part of the van-Hove function. Non-Gaussianity results owing to the enhanced binding events, and the sticky motion of the tracer along with the chain with increasing stickiness. On the other hand, with increasing activity, initially non-Gaussianity increases as the tracer moves through the heterogenous polymeric environment but for higher activity, escapes resulting negative NGP. For higher values of stickiness, the trapping time distributions of the passive tracer particle broadens and have long tails. On the other hand, for a given stickiness with increasing self-propulsion force, the same becomes narrower and have short tails. We believe that our current simulation study will be helpful in elucidating the complex motion of activity-driven probes in viscoelastic media.
\end{abstract}


\maketitle


\section{Introduction}
 \noindent The transport of macromolecules and nanoprobes in  biological medium during biological functions, as well as targeted drug delivery is mainly governed by two mechanisms, either imparted by simple thermal fluctuations, also known as the passive transport, or by {a self-propelled motion generated by burning chemical fuels}. Thermal fluctuations produce random movement, whereas the propulsion force provides a directed motion, also known as active transport. Typical passive transports occur through nuclear pore complex \cite{fragasso2021designer, herrmann2009near, chakrabarti2014diffusion}, anomalous diffusion of telomeres in
the nucleus of mammalian cells, mucus membrane and ete. \cite{ribbeck2017} {Here importance of the repulsive or attractive interaction of the probe with the environment is prominent, which has been experimentally studied.\cite{lieleg2009selective}} {On the other hand, active transport plays, a significant role in the living cells. For example, chemically fueled biomolecules such as microtubules,\cite{sumino2012large} active filaments, molecular motors,\cite{article} etc. These biomolecules self-propel by consuming energy produced from ATP hydrolysis during various biochemical processes and play important roles in intracellular transport, cell motility,\cite{chelakkot2014flagellar} and cell division.\cite{PECREAUX20062111}} These biologically relevant transport phenomena have motivated {a group of} scientists to synthesize and engineer artificial micron and nano-sized particles, capable of self-propulsion, such as Janus particles,\cite{patra2013intelligent, theeyancheri2020translational} chiral particles,\cite{ghosh2009controlled} and vesicles.\cite{joseph2017chemotactic}\\

\noindent Recently, a number of computational studies have revealed the importance of interactions between {the} active particles and their environment, which is one of the causes {for} the highly complex nature of the dynamics. For example, the effective diffusion of the active tracer particle (symmetric, asymmetric) in a crowded media depends not only on the activity but also on the extent of crowding and the interaction between the tracer and the crowders. {The microscopic picture is as follows:} while moving through a crowded and confined space, the tracer particle encounters collisions, and {feels} a range of interactions with the environment, which profoundly influence its dynamics. \cite{theeyancheri2020translational,hou2019,kumar2022chemically,sahoo2022transport,kaiser2020directing,zhao2022,klett2021non,D2CP04253C}  Therefore, the persistent motion of the active tracer is get influenced by the media. {On the other hand, the dynamics of passive probe, such as colloidal particles, polymer chains becomes directed in a bath made of active particles}, but also get affected due to crowding.\cite{chaki2019enhanced,shin2017elasticity}\\

\noindent {One important aspect of the crowded medium is the presence of the specific binding} {sites}. This is important in many {biologically relevent phenomena} such as protein-ligand binding, where ligands bind {to} the specific zone of protein {\textit{via} short-range interactions (hydrogen bond, van der Waals potential), Other examples include enzymatic reactions,\cite{kuchler2016enzymatic,souza2020protein,stayton1995control} signal transduction, immunoreaction, and gene regulation,  as well as} {in case of} targeted drug delivery. {The presence of binding zone is important in protein-ligand binding events in biological processes. Here, the size and topology of the ligands\cite{li2009coarse} and the strength of interaction between the ligand and protein influence the binding-unbinding events. Locating the specific sites on proteins or target molecules is a challenging task,\cite{tsai2008allostery} as the proteins fluctuate. When the binder is chemically fueled, dynamics becomes even more complex. Thus binding--unbinding dynamics of active particle to a macromolecule is worth investigating. Active transport facilitates the faster and efficient delivery of cargoes moving through biological fluids in three dimensions to the specific sites,\cite{chen2019magnetic,liu2012simultaneous}} as well as assist the enzymatic reactions.\cite{gentile2020chemically,feng2020enhanced,jee2019enhanced,ghosh2021enzymes,jee2018catalytic} Recently several experimental\cite{shen2016decorating,gocheva2019look,somasundar2021chemically} and computational studies\cite{gocheva2019look,li2009coarse} focused on elucidating the mechanism involved. {However, experiments on the transport of activity driven\cite{chen2019magnetic,liu2012simultaneous,patra2013intelligent} nano or micro-sized agents in the polymeric media or much less in number. Same is true for computational studies on this topic. As of now, a general understanding of activity-driven transport in viscoelastic media is limited. And therefore, computer simulations can contribute immensely to a deeper understanding of active transport in the viscoelastic crowders. Even simple computer models have potential to add to the current understanding of the subject.} \\

\noindent In the present work, we explore the dynamics of a tracer particle in a crowded medium created by a long polymer chain in three dimensions. We consider two cases, case 1: where all the polymer beads are either attractive or repulsive to the tracer particle, and case 2: where the polymer beads of the central zone are attractive and {the beads of} terminal zones are repulsive to the tracer particle. The self-propelled tracer particle is modeled as an Active Brownian Particle (ABP), which is spherical, {driven by a body fixed force of constant magnitude (F)}. Langevin simulations are performed to elucidate the dynamics of the tracer particle. We examine the dynamics of the tracer particle for different parameters, viz. self-propulsion force (F), size ($\sigma_{\text{tracer}}$), and the interaction strength ($\epsilon$) between the tracer particle and the polymer beads. We observe that in the case of non-zero self-propulsion, the dynamics of the tracer particle exhibits three-step growth with intermediate superdiffusive behavior and faster dynamics compared to the passive tracer with no self-propulsion. For the small self-propulsion force, when all the beads of the polymer are sticky to the tracer, the dynamics of the self-propelled tracer particle shows subdiffusive behavior at the initial time, but as time progresses, self-propulsion dominates over stickiness and superdiffusive behavior appears at the intermediate time. On the contrary, in the case of high self-propulsion force, the subdiffusive region is absent. The dynamics of the self-propelled tracer particle deviates from the Gaussian behavior in the presence of self-propulsion. In case 2, we introduce a specific binding zone on the polymer chain to study the effect of transport of tracer through the crowded medium with specific sites where a fraction of beads forming the central zone of the polymer chain is attractive to the tracer and {other beads} are repulsive. Our simulations show that the tendency of the tracer particle to remain in the trapped states increases with the increase in the stickiness of the tracer particle to the polymer beads. {For a smaller value of self-propulsion force,  tracer particle cease to escape from the trapping zone. However, tendency to escape increases with the self-propulsion. This finding can be used as an efficient method for designing more efficient targeted delivery vehicles which follows faster dynamics, binds to target location and resides there till their action is completed. Our analyses reveal that this is possible with moderate activity and stickiness. On the other hand, for high self-propulsion force, activity dominates over stickiness and the tracer particle easily escapes from the trapping zone} and explore the non-sticky region of the medium. For bigger tracers, the conformation of the trapping zone is decided by the extent of stickiness. Higher the stickiness, more collapsed the trapping zone is. On the other hand, for the smaller tracers, the trapping zone is {disperse} and extended. {In general, studies} on the dynamics of active tracers in the vicinity of polymers with trapping zones are largely lacking in the literature with only few computational studies of active tracer in polymer environment.\cite{zhao2022,D2CP04253C,kim2022active} {Therefore, our investigations are timely and important}. {We strongly believe that our studies shed light on the fantastic interplay between the particle activity and the local structure in a complex heterogeneous medium and microscopic picture of this complex transport process}.\\

\noindent This paper is organized as follows. In Section 2, we describe the model and simulation method. Results and discussion is presented in Section 3 followed by the conclusion in Section 4.\\

\section{Model}
\noindent {Our system is made of a long polymer chain and one tracer particle of diameter $\sigma\textsubscript{tracer}$ (it varies in the range $\sigma, 1.5\sigma, 2\sigma, 4\sigma$). The polymer chain is made of N number of beads of diameter $\sigma_{\text bead}$. The Lennard-Jones parameters $\sigma$, $\epsilon$, and m are the fundamental units of length, energy, and mass respectively. Therefore, the unit of time $(\tau)$ is $\sqrt{m\sigma^2/\epsilon}$. The volume of simulation box is ($12\sigma \times 12\sigma \times 12\sigma$). We carry out independent simulations with polymer of different length (N = 200, 400, 600, corresponding densities are $\phi = 0.06, 0.12, 0.18$ ), where periodic boundery condition are imposed in all directions. The neighboring beads of the polymer are connected by a harmonic potential:} 
\begin{equation}
V\textsubscript{Harmonic}=\frac{1}{2}K(r-r_{0})^2\\
\end{equation}
with the equilibrium bond distance, $r_{0}=0.5$ and force constant, K = 7.0.
\begin{figure*}[!t]
\centering
  \includegraphics[width=\textwidth,keepaspectratio]{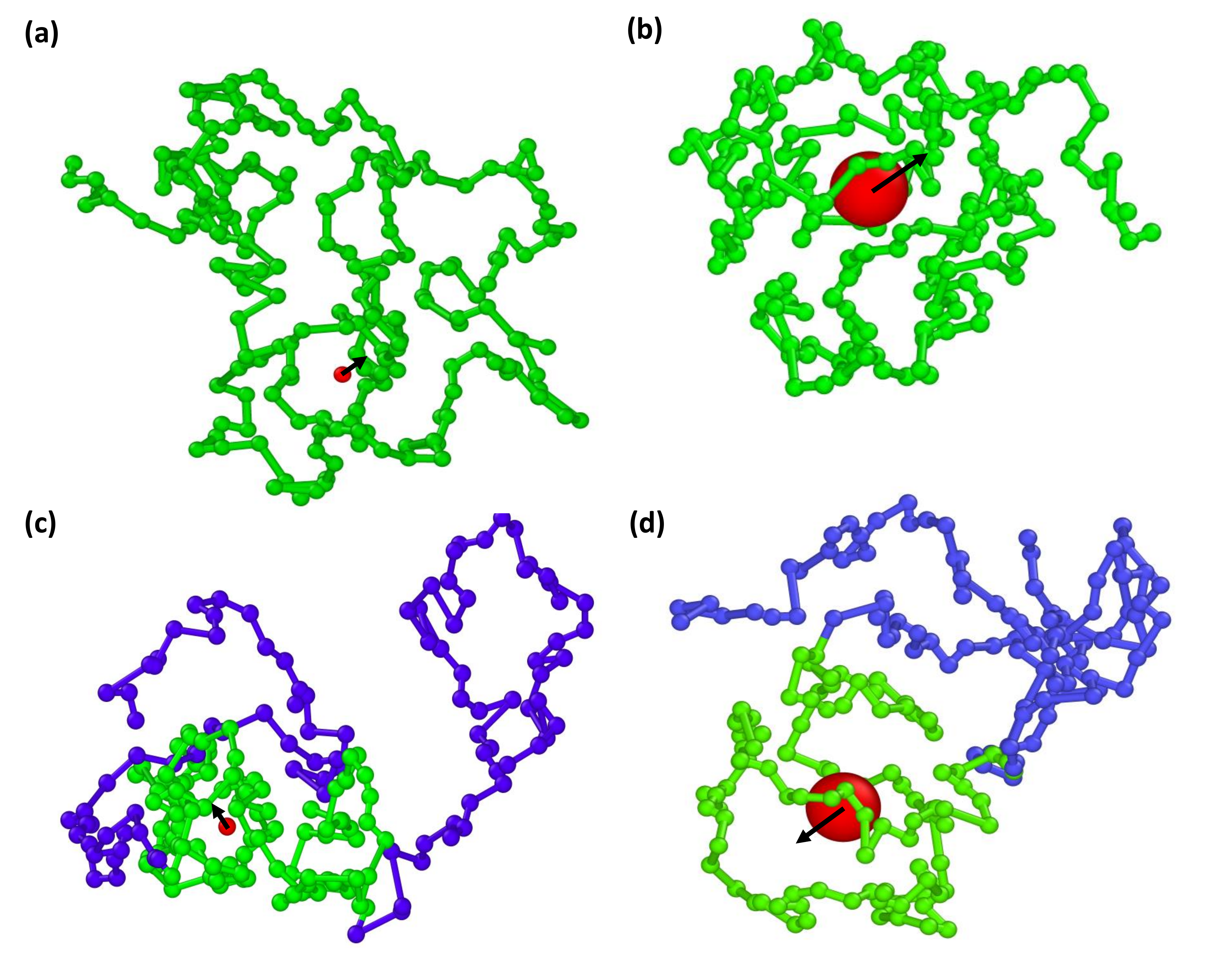}
  \caption{A typical snapshot of the tracer (red) inside the polymeric medium where all the beads are attractive (green) to the tracer for (a) $\sigma\textsubscript{tracer} = \sigma_{\text{bead}}$, (b) $\sigma\textsubscript{tracer} = 4\sigma_{\text{bead}}$, the polymer beads with attractive (green) and repulsive (blue) interactions to the tracer (c) $\sigma\textsubscript{tracer} = \sigma_{\text{bead}}$ and (d) $\sigma\textsubscript{traer} = 4\sigma_{\text{bead}}$. Black arrows represent the instantaneous direction of self-propulsion.}
  \label{fgr:example}
\end{figure*}
The beads of the polymer chain interact by a purely repulsive Weeks–Chandler–Andersen (WCA) potential:\cite{weeks1971role}
\begin{equation}
V\textsubscript{WCA}= 
\begin{cases}
    4\epsilon_{ij}\left[(\frac{\sigma_{ij}}{r_{ij}})^{12}-(\frac{\sigma_{ij}}{r_{ij}})^6\right]+\epsilon_{ij},& \text{if } r_{ij}\leq (2^{1/6})\sigma_{ij}\\
    0,              & \text{otherwise}
\end{cases}
\end{equation}
where $r_{ij}$ is the separation between the {$i^{\text th}$ and $j^{\text th}$ particle}, $\epsilon_{ij}$ is the strength of the interaction with an effective interaction diameter $\sigma_{ij}  = \frac{\sigma_{i} + \sigma_{j}}{2}$. As mentioned in introduction, we consider two cases. For case 1, the tracer particle interacts via the standard attractive Lennard-Jones potential or by the repulsive WCA potential with polymer beads, and the case 2, where beads of the central zone are attractive to the tracer and others are repulsive.  
\begin{equation}
V\textsubscript{LJ}= 
\begin{cases}
4\epsilon_{ij}\left[(\frac{\sigma_{ij}}{r_{ij}})^{12}-(\frac{\sigma_{ij}}{r_{ij}})^6\right],& \text{if } r_{ij}\leq r\textsubscript{cut}\\
0,              & \text{otherwise}
\end{cases}
\end{equation}
 the Lennard-Jones cutoff length $r\textsubscript{cut}=2.5\sigma$. We have varied the interaction strength $\epsilon_{ij}$ and size of the tracer particle $\sigma\textsubscript{tracer}$ in our simulations and fixed the size of beads $\sigma\textsubscript{bead} = \sigma$ of the polymer.
 
\noindent We have implemented the following Langevin equation to simulate the dynamics of a particle ${(i^{\text{th}})}$ with mass $m$ and position r\textsubscript{i}(t) at time t with other particles in the system at position r\textsubscript{j}.

\begin{equation}
 m\frac{d^{2}\textbf{r}_{i}(t)}{dt^{2}}=-\zeta\frac{d\textbf{r}_{i}}{dt}-\Sigma_{j}\nabla V(\textbf{r}_{i}-\textbf{r}_{j})+\textbf{f}_{i}(t)+F\textbf{{n}}
 \end{equation}
 
 \begin{equation}
 \frac{d\textbf{n}}{dt}=\pmb{\eta}(t)\times \textbf{n}
 \label{eq:example}
 \end{equation}
 where $V(r) = V\textsubscript{LJ} + V\textsubscript{WCA} + V\textsubscript{Harmonic}$ is the resultant pair potential between the $i^{\text {th}}$ and $j^{\text {th}}$ particles. $V\textsubscript{LJ}=0$ for purely repulsive interaction and $V\textsubscript{WCA}=0$ for purely attractive interaction. Here we consider high friction coefficient $\zeta=1000$ \text{(damping factor} = $0.001, \zeta=\frac{m}{damp})$, therefore the dynamics is {practically} overdamped. Thermal fluctuation is taken into account by the Gaussian random force $f_{i}(t)$, following the fluctuation-dissipation theorem
 \begin{equation}
 \langle f_{i}(t)\rangle = 0
 \end{equation}
 
 \begin{equation}
 \langle f_{\alpha}(t_{1}) f_{\beta}(t_{2})\rangle=6\zeta k_{B}T\delta_{\alpha \beta}\delta(t_{1}-t_{2})
 \end{equation}
where $k_{B}$ is the Boltzmann constant, T is the temperature and $\alpha$ and $\beta$ represent Cartesian components. {The activity is modeled as $F\bf{n}$, where F represents the magnitude of the active force with the orientation specified by the unit vector $\bf{n}$ which changes according to eq.~\ref{eq:example}. $\pmb{\eta}(t)$ is the stochastic vector, which follows Gaussian distribution with $\langle \pmb{\eta} (t)\rangle=0$ and the time correlation is given by $\langle\pmb{\eta}_{\alpha} (t_{1})\pmb{\eta}_{\beta} (t_{2})\rangle = 2D_{R}\delta_{\alpha\beta}\delta(t_{1}-t_{2})$, {where $D_{R} = \frac{1}{2\tau_{R}}$} \cite{hou2019} and $\tau_{\text{R}}$ is the persistence time. The unit vector \textbf{n} can be expressed in the form of spherical polar coordinates ($\sin\theta \cos\phi, \sin\theta \sin\phi, \cos\theta$). 
\begin{equation}
 \frac{d\theta}{dt} = \eta_{y} \cos\phi - \eta_{x} \sin\phi
 \end{equation}
 \begin{equation}
 \frac{d\phi}{dt} = \eta_{z} - \eta_{x}\frac{\cos\theta}{\sin\theta}\cos\phi - \eta_{y}\frac{\cos\theta}{\sin\theta}\sin\phi,
 \end{equation}
 \noindent where $\eta_{x}$, $\eta_{y}$ and $\eta_{z}$ are component of the Gaussian white noise $\pmb{\eta}$ in Cartesian coordinate space.
F=0 for the polymer beads as well as when the tracer is passive and non-zero for self-propelled tracer particle.}\\

\noindent {All the simulations are done} using LAMMPS.\cite{plimpton1995fast} Initially, the system is equilibrated by simulating for $10^{7}$ steps. For each simulation, the time step is chosen to be $5\times 10^{-4}\tau$. All simulations are carried out in the presence of a Langevin thermostat and the equation of motion is integrated using the Velocity Verlet algorithm in each time step. Final simulations are carried out for $10^8$ steps. The positions and velocities of the tracer particles are recorded every $100^\text{th}$ steps.\\

\section{Results and discussion}
\noindent {In this section we discuss and explain the results in detail.}
{\subsection{Homopolymeric Medium}
\noindent In this section, we consider the case when the polymer chain is made of chemically identical beads either all repulsive or all are attractive to the tracer. These beads are mutually repulsive (WCA).}\\

\begin{figure*}
\centering
  \includegraphics[width=\textwidth,keepaspectratio]{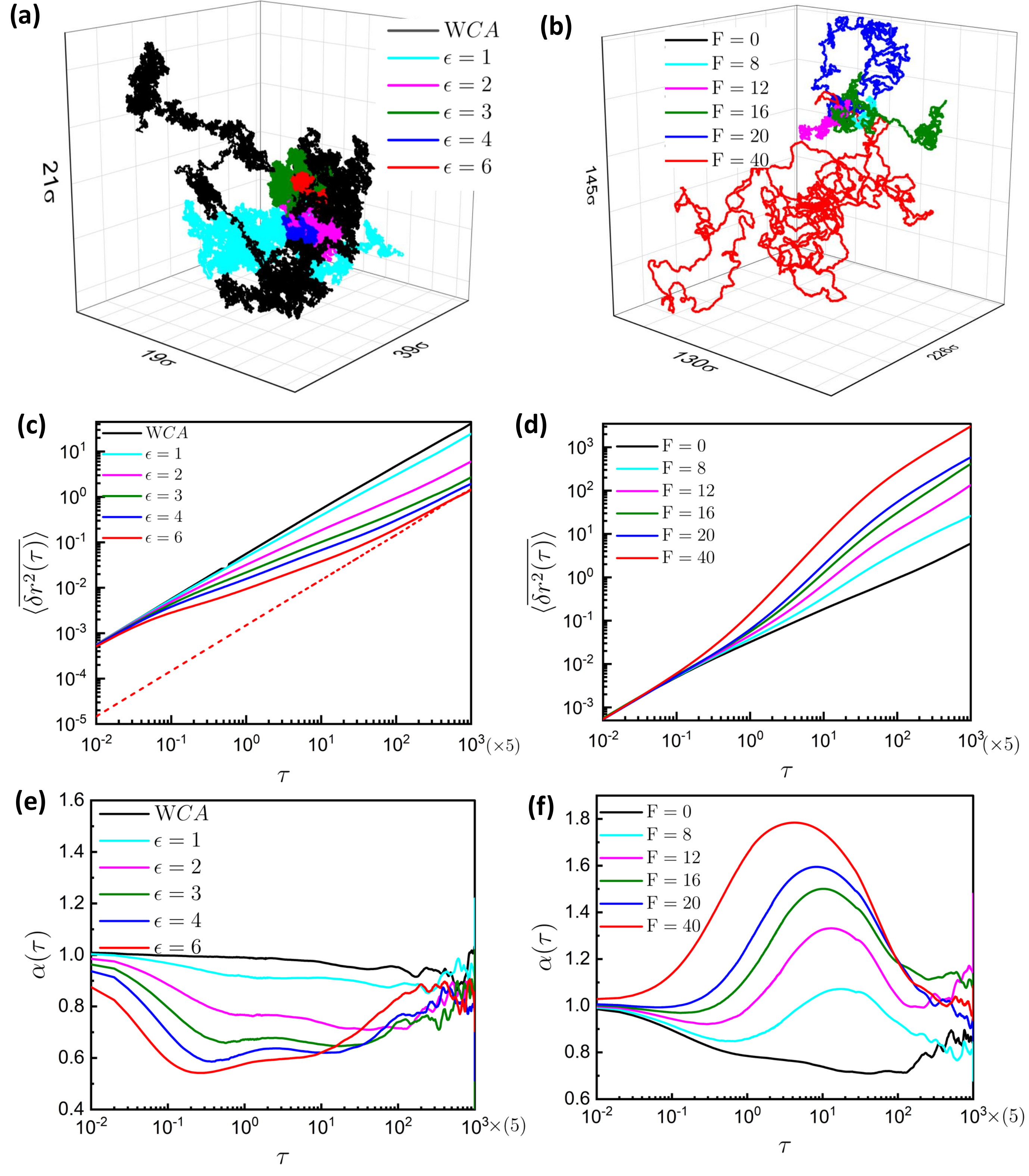}
  \caption{Trajectories of tracer particles for (a) different $\epsilon$ at F=0,  (b) different F at $\epsilon= 2$, log-log plot of $\langle\overline{\delta r^2(\tau)}\rangle$ vs $\tau$ for (c) different $\epsilon$ at F=0, dashed one is $\langle\overline{\delta r^{2}_{c}(\tau)}\rangle\times R_{g}$ (d) different F at $\epsilon=2$, log-linear plot of $\alpha(\tau)$ vs $\tau$ for (e) different $\epsilon$ at F=0, and (f) different F at $\epsilon=2$ (N 400, $\sigma\textsubscript{tracer} = \sigma$).}
  \label{fgr:msd}
\end{figure*}
\subsubsection{Mean Square Displacement, MSD ($\langle\overline{\delta r^2(\tau)}\rangle$) and Time exponent $(\alpha(\tau))$}
\noindent In order to gain insights into the dynamics of the tracer particle, we calculate time-averaged MSD, $\overline{\delta r_{i}^2(\tau)}=\frac{1}{T\textsubscript{max}-\tau}\ \int_{0}^{T\textsubscript{max}-\tau} [r_{i}(t+\tau)-r_{i}(t)]^2 \,dt$, from the time progression of r(t), where T\textsubscript{max} is the total run time and $\tau$ is the lag time. Next, to find time and ensemble-averaged MSD we calculate the average of N independent trajectories $\langle\overline{\delta r^2(\tau)}\rangle=\frac{1}{N}\sum\limits_{i=1}^N \delta \overline{r\textsubscript{i}^2(\tau)}$. {In case of normal diffusion $\langle\overline{\delta r^2(\tau)}\rangle$ grows linearly with $\tau$, but if the dynamics is anomalous,\cite{jain2016diffusing,Miyaguchi2022} then $\langle\overline{\delta r^2(\tau)}\rangle \propto \tau^{\alpha}$ with $\alpha < 1$ (subdiffussion)\cite{kumar2019transport, ghosh2015non,Miyaguchi2022,jee2019enhanced} or $\alpha > 1$ (superdiffusion). Therefore $\alpha(\tau)=\frac{d \text{log}(\langle\overline{\delta r^2(\tau))}\rangle}{d \text{log}(\tau)}$, gives this running exponent. }\\
 
 \noindent{To elucidate} the effect of crowding, {stickiness of the tracer particle} with {the} polymer beads, and self-propulsion force (F) on the dynamics of the tracer particle, we calculate the $\langle\overline{\delta r^2(\tau)}\rangle$ and $\alpha(\tau)$. {first we   we focus on the effect of strength of attraction ($\epsilon$) between the tracer particle and polymer beads for the passive (F = 0). These observations reveal that in the case of purely repulsive (WCA) potential, the dynamics of the tracer particle is faster than that in case of the attractive potential (Fig.~\ref{fgr:msd}(a and c)).} On the other hand, when the tracer particle is attractive to the polymer beads, it gets locally trapped $(\text{Movie}\_\text{S1})$ inside the polymer which slows the dynamics and results in subdiffusion (Fig.~\ref{fgr:msd}c) at intermediate time. This effect is more pronounced with increasing $\epsilon$ which makes the dynamics of the tracer particle { strongly} subdiffusive, as shown in Fig.~\ref{fgr:msd}c. Further, evolution of  $\alpha(\tau)$ with time supports similar behavior with $\alpha=1$ for normal diffusion, and $\alpha<1$ for subdiffusive behavior as shown in (Fig.~\ref{fgr:msd}e). The subdiffusive behavior at the intermediate time is due to the local trapping of the tracer particle as the polymer interacts strongly with the tracer { and slows down the} dynamics. When the polymer changes its conformation, particle escapes from these local traps. {In the case of high $\epsilon$, the passive tracer approach the diffusive limit at an earlier time as compared to {the one with} smaller $\epsilon$. This is due to the slow movement of tracer particle in the vicinity of polymer as the tracer is moving {along} with the polymer at very high values of stickiness $\epsilon$. This is further verified by comparing the re-scaled center of mass $\langle\overline{\delta r^{2}_{c}(\tau)}\rangle$ ($\langle\overline{\delta r^{2}_{c}(\tau)}\rangle\times R_{g}$) of the central binding zone beads of polymer with $\langle\overline{\delta r^2(\tau)}\rangle$ of the tracer in progress of time, here we observe that $\langle\overline{\delta r^{2}_{c}(\tau)}\rangle$ merges with  $\langle\overline{\delta r^2(\tau)}\rangle$ in the long time limit.} Next, to understand the effect of self-propulsion we take the self-propelled tracer particle with $\epsilon=2$, and vary the self-propulsion force. We observe that, the trajectories of the self-propelled tracer particle spread with increase in self-propulsion force (Fig.~\ref{fgr:msd}b) {this} {in addition,}  $\langle\overline{\delta r^2(\tau)}\rangle$ grows in three steps, normal diffusion $(\alpha(\tau)=1)$ at short time, superdiffusion $(\alpha(\tau)>1)$ at the intermediate time followed by enhanced diffusion at the longer time, and the dynamics is faster in comparison to the passive particle (Movie\_S1 and Movie\_S2) as shown in Fig.~\ref{fgr:msd}d and Fig.~\ref{fgr:msd}f. Also {on} increasing F, the long time value of $\langle\overline{\delta r^2(\tau)}\rangle$ increases (Fig.~\ref{fgr:msd}d). We find that self-propelled tracer particle {even} shows subdiffusive behavior (Fig.~\ref{fgr:msd}f) for small $F (F=8)$ at a short time, but as time progresses self-propulsion force overcomes the stickiness and crowding (Movie\_S2), and as a result, superdiffusive behavior is observed at the intermediate time. But in the case of higher self-propulsion force, subdiffusive region in the short time is absent, because the F dominates overcrowding and stickiness \cite{theeyancheri2022silico} in a shorter time (Movie\_S3). {in the long time limit}, the direction of the self-propelled particle is randomized and behaves like a diffusive particle with high value of diffusivity.\\
 
 \noindent {Further to check the effect of polymer density on the tracer dynamics we perform simulations for different {densities} of the {polymer} by changing the number of beads of the polymer (N = 200, 400, 600, corresponding densities are $\phi = 0.06, 0.12, 0.18$) for a fixed box size$12\sigma \times 12\sigma \times 12\sigma$. The dynamics shows qualitatively similar trend at different densities of polymer for different activity and $\epsilon$, which is shown in Fig.~S2. Only difference is that dynamics of the tracer slows down with increase in density as shown in Fig.~S3(a, b). This effect is negligibly small for high attraction strength of the tracer to the polymer beads, due to localised motion of the tracer.  In case of passive tracer particle MSD for all densities merge for $\epsilon = 2$, where motion is localized. However on introducing self-propulsion, activity dominates over stickiness and a clear difference in dynamics is observed for different densities.\\}
 
\noindent Next we calculate the $\langle\overline{\delta r^2(\tau)}\rangle$ (Fig.~S4(a)) and $\alpha(\tau)$ (Fig.~S4(b)) to analyze the dynamics of the tracer particle {of different sizes} ($\sigma\textsubscript{tracer} = \sigma, 1.5\sigma, 2\sigma$), keeping $\epsilon=2$ and $F=60$. The results reveal that increasing the size of the tracer particle, dynamics slows down because the bigger {sized} tracer particle interacts more strongly with the polymer, {as it interacts with several monomers due to the larger size}. Apart from this, we notice that {for the bigger sized tracer particle,} the transition from normal diffusion to superdiffusion and superdiffusion to enhanced diffusion is delayed. These observations are related to the rotational dynamics of the tracer particle, in the form of persistence time $(\tau\textsubscript{R})$. For $\tau<\tau\textsubscript{R}$, tracer particle follows normal diffusion, for $\tau\approx\tau\textsubscript{R}$ particle follows superdiffusive dynamics, and finally for $\tau>\tau\textsubscript{R}$ again tracer follows normal diffusion with enhanced diffusion coefficient. $\tau\textsubscript{R}$ is related to the size of the tracer particle as $\tau\textsubscript{R}\propto R^3$ where R is the radius of the tracer particle. Hence for {larger} tracer particle, {$\tau\textsubscript{R}$is higher and the transition is delayed}.\\
\begin{figure*}
\centering
  \includegraphics[width=\textwidth,keepaspectratio]{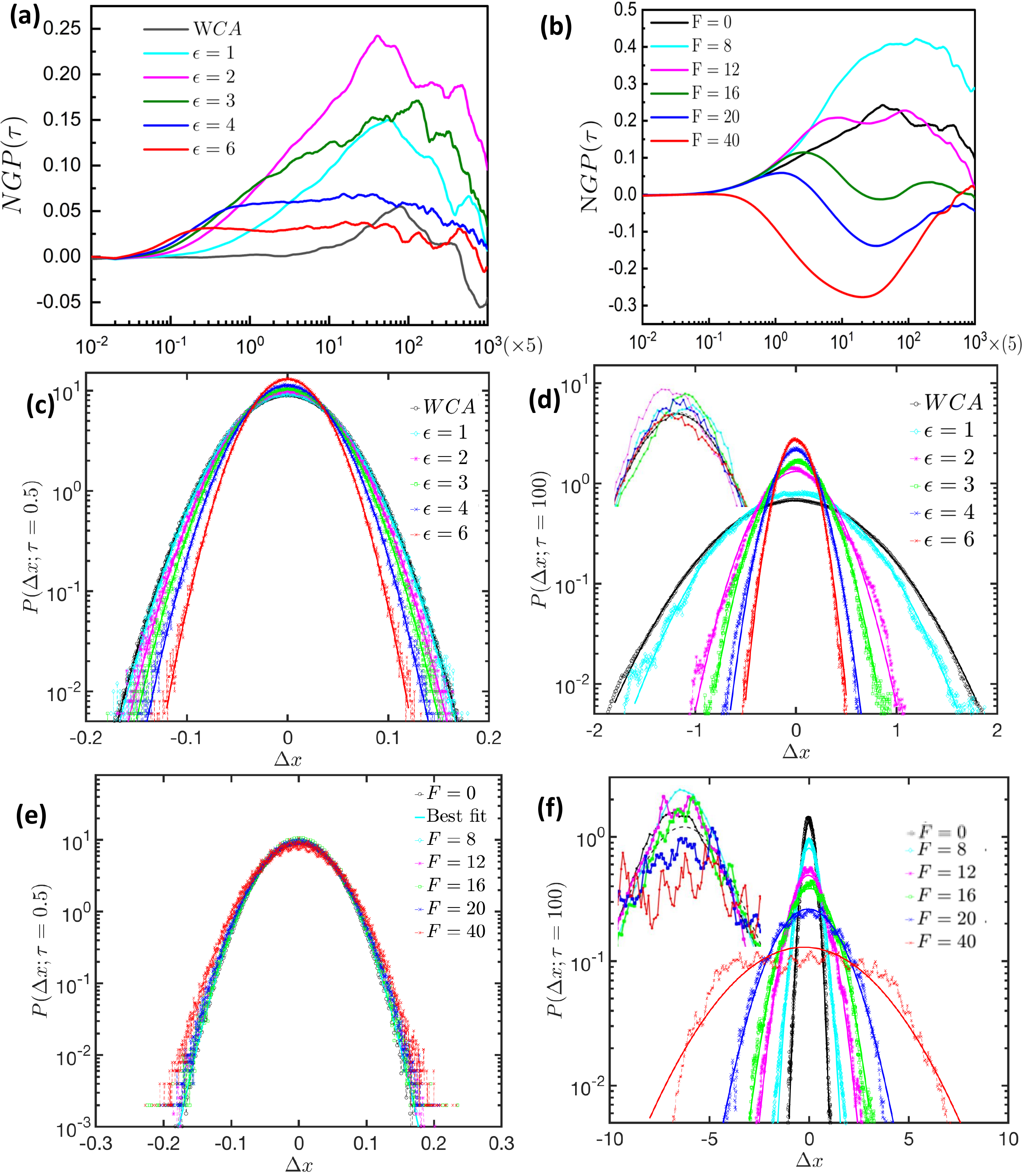}

  \caption{Log-linear plot of $\text{NGP}(\tau)$ vs $\tau$ for (a) different $\epsilon$ and $F=0$, (b) different F ($\epsilon=2$), plot  of $P(\Delta x, \tau = 0.5)$ for (c) different $\epsilon$ (F = 0), (d) different F ($\epsilon=2$), and $P(\Delta x, \tau = 100)$ for (e) different $\epsilon$ (F = 0) and (f) different F ($\epsilon =2$). Solid line in figure (c) to (f) is the best Gaussian fit for respective distribution (N = 400, $\sigma\textsubscript{tracer} = \sigma$). Insets show the enlarged view of the peak of the rescaled plots of the respective van-Hove distributions. }
  \label{fgr:ngp_vanHove}
\end{figure*}
\begin{figure*}
\centering
  \includegraphics[width=\textwidth,keepaspectratio]{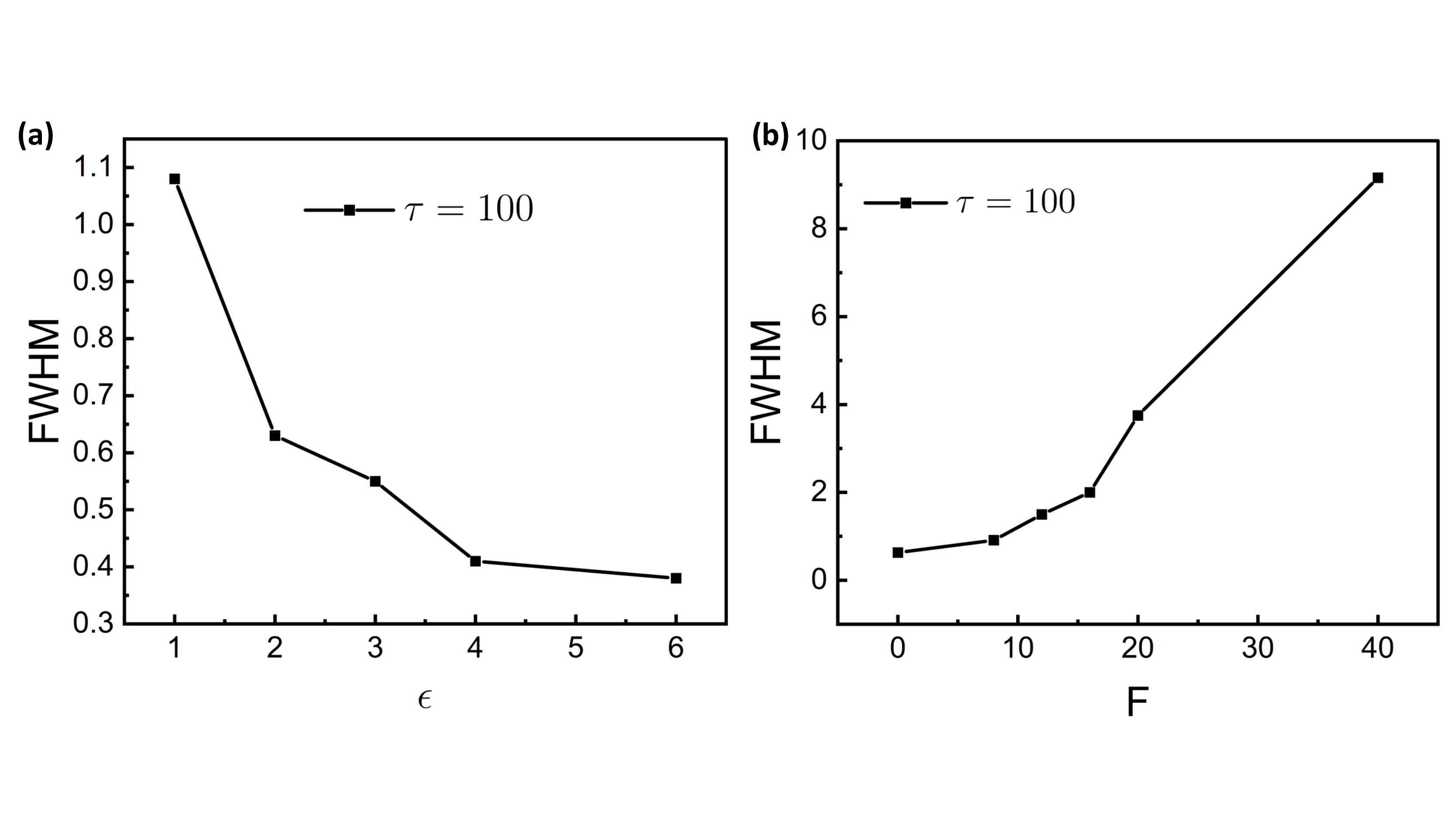}
  
   \caption{Plots for full width half maxima (FWHM) of the van-Hove functions against (a) $\epsilon$ (F = 0) and (b) F ($\epsilon = 2$) (N = 400, $\sigma\textsubscript{tracer} = \sigma$).}
  \label{fgr:fwhm}
\end{figure*}

\subsubsection{Non-Gaussian Parameter $(\text{NGP} (\tau))$ and Probability Distribution of displacement}
\noindent {To have a {deeper} picture of the dynamics}, we calculate the NGP and self part of the van-Hove correlation function (probability distribution function {of displacement}) for the tracer. In general, these two parameters are used to describe the non-Gaussianity in the dynamics, where the NGP in three dimensions is defined as:\cite{hofling2013anomalous}

\begin{equation}
 NGP(\tau)=\frac{3}{5}\frac{\langle\overline{\delta r^4(\tau)}\rangle}{\langle\overline{\delta r^2(\tau)}\rangle^2}-1
 \end{equation}

\noindent For normal diffusion which follows Gaussian distribution, $\text{NGP}(\tau)$ = 0, while the parameter $\text{NGP}(\tau)$ becomes non-zero for a process with non-Gaussian distributions. Example of non-Gaussian distribution includes subdiffusive continuous time random walk (CTRWs) or heterogeneous diffusion processes\cite{cherstvy2021inertia,ernst2014probing,dueby2022size,acharya2017fickian} or dynamics of self-propelled tracer particle.\cite{kurzthaler2016intermediate, goswami2022motion}\\

\noindent {In case of a passive tracer for a given stickiness ($\epsilon)$}, $\text{NGP}(\tau)$ {shows following behavior with the progress of time (Fig.~\ref{fgr:ngp_vanHove}a)}. { The short time dynamics is Gaussian (NGP = 0) and starts to increase from zero at intermediate time as a function of increasing $\epsilon$. Maximum non-Gaussianity is observed at $\epsilon = 2$ and further increasing $\epsilon$, $\text{NGP}(\tau)$ starts to {reduce}. This non-monotonicity in $\text{NGP}(\tau)$ is attributed to the slower dynamics of tracer at higher $\epsilon$, tracer-polymer interaction dominates and hence the tracer is failed to explore the heterogeneity of the medium. In contrast, for smaller $\epsilon$ values the tracer smoothly explores different parts of the medium and undergoes binding-unbinding events\cite{fernandez2020diffusion,acharya2017fickian,guha2022multivalent} resulting more pronounced non-Gaussianity. Next, we check how the non-Gaussianity {shows up} in the dynamics of active tracer for a fixed values of stickiness $\epsilon = 2$. We notice that dynamics is Gaussian at short time, starts deviating from Gaussianity at intermediate time. $\text{NGP}(\tau)$ increases with F until a moderate activity (F=8), after which it decreases and {even} becomes negative for higher F. The self-propulsion {is} not sufficient to overcome the sticky interaction, the tracer more frequently collides and slowly approach the diffusive limit with the help of polymer dynamics. On the other hand, for larger F, the activity dominates over {sticky} interaction and {the tracer hits} the diffusive limit faster compared to lower values of F. For very high F, the tracer displays prolonged persistent motion. This can be explained as follows: for higher activity, the mean square displacements of the tracer become larger, this is due to higher effective temperature, {that results in the broadening of the van-Hove distributions}.} On the other hand, the directed motion gives, that all displacement will be equally probable, and  {appears} as platue in van-Hove distribution of the displacement (defined below) is as shown in Fig.~\ref{fgr:ngp_vanHove}f, where tail drops sharply for larger displacements in comparison to the normal distribution. Thus $5{\langle\overline{\delta r^2(\tau)}\rangle^2}$ dominates over $3{\langle\overline{\delta r^4(\tau)}\rangle}$ Fig.~S5.\\

\noindent The self part of the van-Hove correlation function (probability distribution function), $ P(\Delta x;  \tau)$ is defined as follows $ P(\Delta x;  \tau) \equiv\langle\delta (\Delta x-(x(t+\tau)-x(t)))\rangle$, where $x(t +\tau)$ and $x(t)$ are the positions of the tracer particle along x-direction at time $(t +\tau)$ and $t$ respectively. {Same can be defined for other two directions as well as for the radial displacement. However, since our system is isotropic, we analyse $ P(\Delta x;  \tau)$. The $P(\Delta x; \tau)$ are plotted with the corresponding Gaussian distribution fits for the normal Brownian motion, $P(\Delta x;  \tau) = \frac{1}{\sqrt{2\pi\langle\Delta x^{2}\rangle}}\exp{\Big(-\frac{\Delta x^{2}}{2\langle\Delta x^{2}\rangle}\Big)}$ for different {values of F and $\epsilon$}. It is evident from Fig.~\ref{fgr:ngp_vanHove}(c and e) that on increasing $\epsilon$, distributions become narrower. On the other hand, on increasing F, {at intermediate time ($\tau=100$) the distributions show interesting behavior, especially sudden drops of the tails at larger F are clearly visible, accounting for the directed motion of the tracer (Fig.~\ref{fgr:ngp_vanHove}d). Whereas, in case of a lower value of $\tau = 0.5$, the deviation from Gaussianity is not present in both the cases for different $\epsilon$ and also for range of F as shown in Fig.~\ref{fgr:ngp_vanHove}(e and f). This is essentially a consequence of small time where effect of external factor is not pronounced, this is also evident from the plot of $\text{NGP}(\tau)$ (Fig.~\ref{fgr:ngp_vanHove}(a and b)}). In order to clearly distinguish the difference between different parameter and lag time, horizontal axis is re-scaled to the square root of the corresponding second moment, $\sqrt{\langle\delta x^2(\tau)\rangle}$.\cite{nahali2018nanoprobe} For freely diffusive particle the distribution of re-scaled displacement, $\boldsymbol{\Delta X} = \frac{\Delta x}{\sqrt{\langle\Delta x^2\rangle}}$ is expected to be describe by the universal Gaussian function with zero mean and unit variance. Hence rescaled distributions for repulsively interacting tracer with {the} polymer {beads} are found out to be universal Gaussian distribution like free particle. The re-scaled $P(\Delta X; \tau)$ are plotted with the universal Gaussian distribution function $P(\Delta X;  \tau) = \frac{1}{\sqrt{2\pi\langle\Delta X^{2}\rangle}}\exp{\Big(-\frac{\Delta X^{2}}{2\langle\Delta X^{2}\rangle}\Big)}$ for different {values of F} and $\epsilon$. At short time $\tau = 0.5$, the deviation from Gaussianity is less pronounced for all values of $\epsilon$ and also for all F as shown in Fig.~S6(a, c). This is consistent with $\text{NGP}(\tau)$ (Fig.~\ref{fgr:ngp_vanHove}(a, b)). It is evident from Fig.~S6(b) and in the inset of (Fig.~\ref{fgr:ngp_vanHove}(d)) that the deviation from the fitting as cusp around $\Delta X = 0$ increase with increasing $\epsilon$ up to $\epsilon = 2$ and starts to decrease with further increasing $\epsilon$ value at $\tau = 100$. On increasing F, at intermediate time, for active tracer we find interesting behavior in that deviation from the fitting as cusp around $\Delta X = 0$ increase up to moderate activity (F = 8), start to drop with further increasing F, and at very high self-propulsion shows a platue at the peak and tails of the distribution suddenly drops. This accounts for the large directed motion of the tracer, which is shown in Fig.~S6(d) and in the inset of (Fig.~\ref{fgr:ngp_vanHove}(f)).} \\

\noindent The changes in the dynamics of tracer particle with F and $\epsilon$ are further investigated by calculating the full-width half maxima (FWHM) of the displacement probability distribution functions as shown in Fig.~\ref{fgr:fwhm}(a and b). The FWHM values decrease on increasing $\epsilon$ in the case of the passive tracer particle in support of the narrower distributions with increasing stickiness as depicted in Fig.~\ref{fgr:ngp_vanHove}(d). On the other hand, in the case of self-propelled tracer particles, FWHM values increase with an increase in the self-propulsion force, F. This is evident from the broader van-Hove distributions with increasing F (Fig.~\ref{fgr:ngp_vanHove}(f)). The appearance of the sudden jump in FWHM between self-propulsion force, $F=8$ and $F=20$ is attributed to the tracer particle escaping from the traps created by the polymer chain. This observation also goes hand in hand with the decrease in the value of the NGP at the intermediate time on increasing the self-propulsion force, $F=8$ to $F=20$, as can be seen in Fig.~\ref{fgr:ngp_vanHove}b

{\subsection{Polymer with a Specific Binding Zone}
\noindent The binding-unbinding events are crucial in biological systems to properly carry out the metabolic activities and their assigned functions\cite{weng2011study,morris2005real}. The binding sites behave as flexible networks. Previous studies attempt to reveal the role of flexibility\cite{james1982conformational,weng2011study,morris2005real,riziotis2022conformational,tabatabaei2011simulational,quesada2022coarse} of the binding pocket and the underlying physics. The strength of binding is also important in proper functioning and it depends on the interaction and conformation of the polymer. Next, we introduce a binding zone in the polymer chain by making the central 200 beads of the polymer to be attractive with the tracer and the terminal sections remain as  repulsive.} 

\begin{figure*}
\centering
  \includegraphics[width=\textwidth,keepaspectratio]{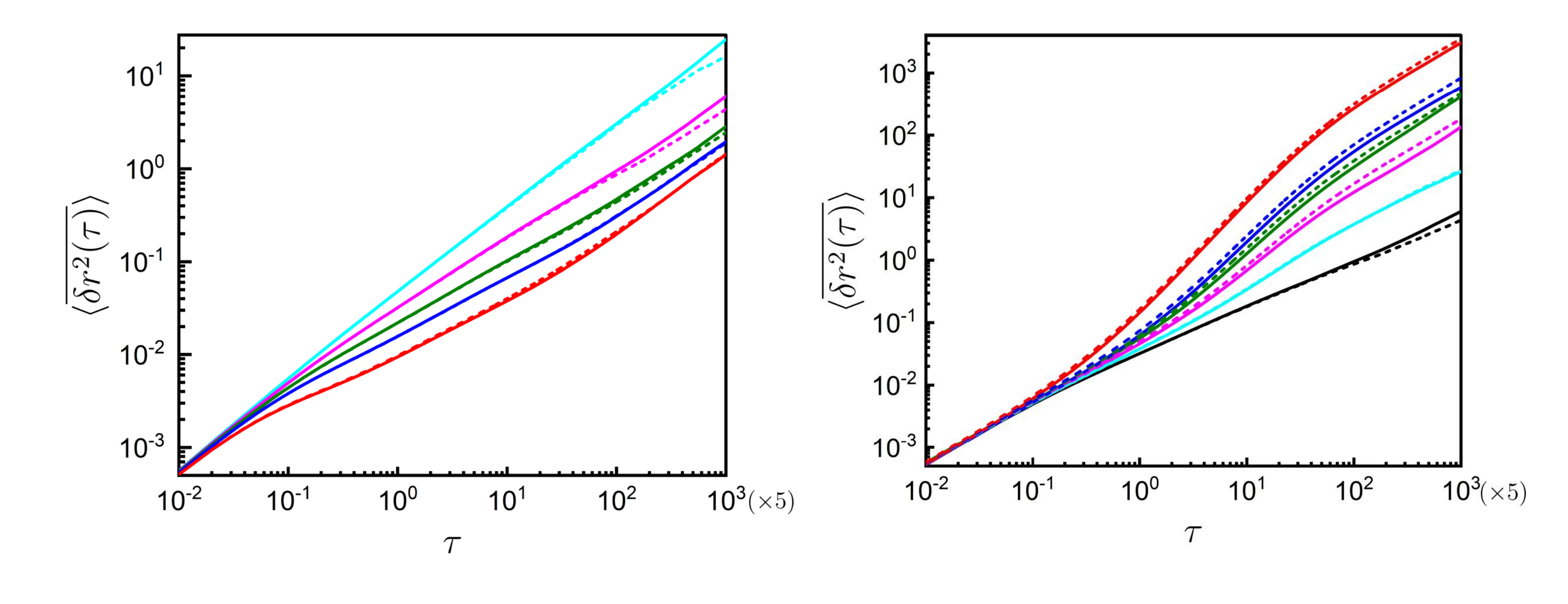}
  
 \caption{Log-log plot for $\langle\overline{\delta r^{2} (\tau)}\rangle$ in progress of time for different (a) $\epsilon$ (F = 0), (b) F ($\epsilon = 2$). solid lines are for polymer having all sticky beads and dashed are for polymer with specific binding zone (N = 400, $\sigma\textsubscript{tracer} = \sigma$).}
  \label{fgr:homo_hetro}
\end{figure*}

{\subsubsection{Dynamics}
\noindent  For the polymer with specific binding zones, we find that the dynamics of the tracer shows similar trend like the polymers with all sticky interaction for increasing $\epsilon$ or activity, which is shown in Fig.~S7. Surprisingly, the dynamics of the tracer is slower for moderate range of $\epsilon$ in the case of polymer with uniform binding zone compared to polymer with central binding pocket (Movie\_S4). For higher $\epsilon$, $\langle\overline{\delta r^{2}(\tau)}\rangle$ curves merge as shown in Fig.~\ref{fgr:homo_hetro}a. This implies that the repulsive terminal part of the polymer force the tracer to stay in the binding pocket at moderate range of stickiness. But for higher $\epsilon$ values, the stickiness dominates and tracer stays inside the binding zones leading to a similar dynamics. When the tracer is active, there is a competition between the interaction and activity. Hence, with increasing F the tracer efficiently moves around and escapes easily from the binding zones leading to a faster dynamics in the case of polymer with specific binding zones than the polymer with all sticky zones (Movie\_S5 and Movie\_S6). This is due to the smaller size of the binding zone in the former case comparing to the later. The difference is pronounced as a function of increasing F as shown in Fig~\ref{fgr:homo_hetro}b.}

\subsubsection{Trapping}
\begin{figure*}[!t]
\centering
  \includegraphics[width=\textwidth,keepaspectratio]{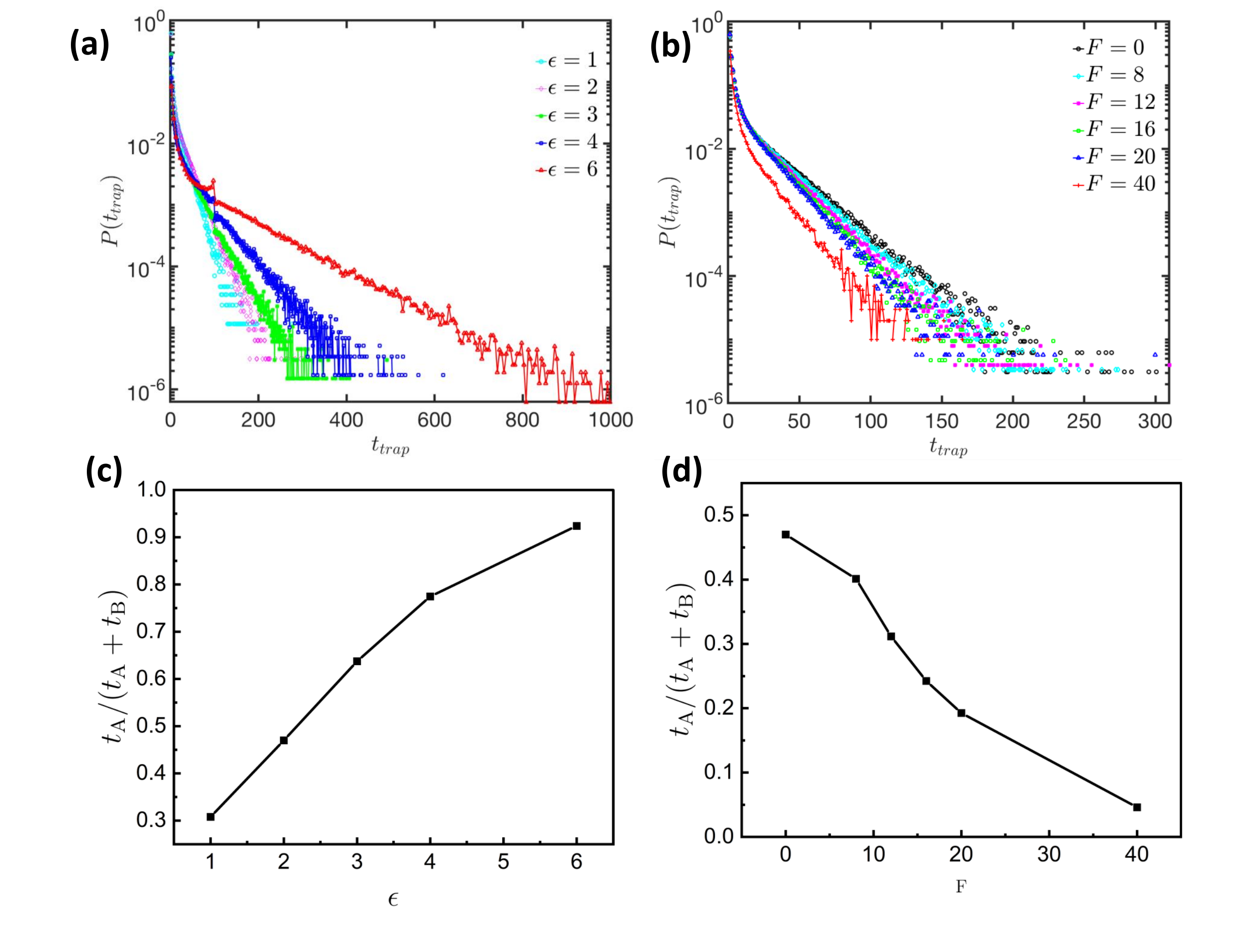}

  \caption{Log-linear plots of $P(t\textsubscript{trap})$ vs $t\textsubscript{trap}$ for mobile polymer with (a) different $\epsilon$ (F = 0), (b) different F $(\epsilon=2)$, ($\frac{t\textsubscript{A}}{t\textsubscript{A}+t\textsubscript{B}}$) with mobile polymer (c) vs $\epsilon$ (F = 0), and (d) vs F $(\epsilon = 2)$ (N = 400, $\sigma\textsubscript{tracer} = \sigma$). } 
  \label{fgr:trap_time}
\end{figure*}

\noindent {In the earlier sections, we find that the tracer stays close to the binding zones in the polymer and this is significantly affect the dynamics. The time tracer spends in the vicinity of binding zones is crucial in effective transport of tracer through polymeric medium. To quantify this time we use the concept of trapping time ($t\textsubscript{trap}$), its distribution and the moments}. $t\textsubscript{trap}$ is defined as follows: if the tracer particle travels $0.15\sigma$ distance or less in a {time span of five times the smallest lag time (i.e, $5\tau$)}, then the tracer particle is considered to be trapped, else it is free.
We consider a polymer chain with a fraction of internal monomers sticky to the tracer, whereas all the other monomers are repulsive to the tracer. Therefore, with a polymer having a centrally distributed trapping zone for the tracer, we calculate the trapping time distribution $P(t\textsubscript{trap})$. The corresponding mean trapping time $\langle t\textsubscript{trap}\rangle$ is the average of overall trapping events, averaged over the trajectories of the tracer particle for a given set of parameters. As one can notice that the tails of the distributions become increasingly longer with increasing $\epsilon$.  This implies that the probability of being in the trapped states becomes significantly larger (Fig~\ref{fgr:trap_time}a) with increasing stickiness ($\epsilon$). {However,} the increase in the value of F leads to shorter tails. This is expected, as with increasing self-propulsion force, the tracer can easily escape from the traps, resulting in much narrower trapping time distributions (Fig~\ref{fgr:trap_time}b). As can be seen from Table 1, the average trapping time drops with the increase in the self-propulsion force.

\begin{table}[h]
\small
  \caption{\ Average trapping time $\langle t\textsubscript{trap}\rangle$  for different $\epsilon$ (F = 0) and F ($\epsilon = 2$), expressed in terms of Lennard Jones unit, {$\sqrt{\frac{m \sigma^2}{\epsilon}}$. }}
  \label{tb2:example2}
  \begin{tabular*}{0.48\textwidth}{@{\extracolsep{\fill}}|c|c|c|c|}
    \hline
    $\epsilon$ & $\langle t\textsubscript{trap}\rangle(\tau)$ & F & $\langle t\textsubscript{trap}\rangle(\tau)$\\
    \hline
    1 & 8.98 & 0 & 11.05 \\
    2 & 11.05 & 8 & 10.23 \\
    3 & 14.40 & 12 & 9.32 \\
    4 & 19.80 & 16 & $ 8.65$ \\
    6 & 43.30 & 40 & $ 6.88$\\
    \hline
  \end{tabular*}
\end{table}
\noindent Subsequently, we calculate the total trapping time (t\textsubscript{A}) experienced by the tracer particle during a trajectory of time-span T. We thus sum all the trapping time experienced by the tracer,

\begin{equation}
 t_{A} = \sum\limits_{i}t\textsubscript{trap,i}
 \end{equation}\\
 and the expedition times in the free space,
 \begin{equation}
 t_{B} = \sum\limits_{i}t\textsubscript{free,i}
 \end{equation}\\
 we have total run time, ${T_{\text{total}}}$=t\textsubscript{A}+t\textsubscript{B} and a new quantity relative trapping time, $\frac {t\textsubscript{A}}{t\textsubscript{A}+t\textsubscript{B}}$. This also shows that on increasing $\epsilon$ trapping tendency increases, where the value of relative trapping time initially increases, and for large $\epsilon$ it saturates to 1. In contrast, on increasing F, the relative trapping time decreases and approaches zero for large F.\\ 
 
\noindent {We are also interested to know how the size of the tracer particle and activity {affect} the conformation of the trapping zone of the polymer. For that} we calculate the radius of gyration $(R_g)$, of the trapping zone. $R_g$ is defined as $\sqrt{\frac{1}{N_{\text{trap}}} \sum_{i=1}^{N_{\text{trap}}} m(r_{i}-r_{\text{com}})^2}$, where $N_{\text{trap}}$ is the total number of monomers of the polymer in trapping zone ($N_{\text{trap}}$=100) and $r_{\text{com}}$ is the centre of mass of the trapping zone, and in Fig.~S8 we plot the distribution of this quantity.  We find that in the case of passive tracer particles, the most probable value of $R_g$ decreases with increasing the size of the tracer. For the smaller tracer particle, the $P(R_g)$ shows a broader distribution, whereas {the distribution is narrower for the larger tracer particle}, accounting for a more compact trapping zone as shown in Movie\_S7. On the other hand, in the case of a self-propelled tracer particle with $F=20$ distribution of $R_g$ follows the same trend as in the case of passive, but in comparison to the passive tracer, $R\textsubscript{g}$ distribution is broader for constant $\sigma_{\text{tracer}}$ Which is seen as more fluctuation of polymer chain in the presence of active tracer particle than the passive (Movie\_S8) . From these observations, one can conclude that with increasing size ($\sigma_{\text{tracer}}$) of the tracer particle, the trapping zone of the polymer becomes smaller as then the polymer chain effectively wraps around the tracer, be it passive (Movie\_S6) or self-propelled. The collapse of the trapping zone is owing to the stronger interactions of the polymer beads with the larger-sized tracer particle. However, for the self-propelled tracer particle, distribution of $R_{g}$ is wider than the passive, because of the faster dynamics of the tracer particle. {This is a reflection of not so efficient trapping of the tracer to the binding zone}.

\section{Conclusions}
\noindent The current work represents an extensive simulation-based investigation of tracer particle dynamics in a polymeric medium. The tracer is active i.e, self-propelled and the dynamics is therefore profoundly controlled by the magnitude of self-propulsion in addition to the size of the tracer, mechanical properties of the polymer chain, and the nature of the interaction between the tracer and the monomers of the chain. When the tracer is repulsive to the polymer chain, or in other words, the interaction between the tracer and the monomers of the chain is modeled by WCA potential, the dynamics is purely diffusive {in case of a passive tracer} but becomes superdiffusive at the intermediate time if the tracer is self-propelled. On the other hand, if the tracer has some sort of sticky interaction with the polymer chain, which is modeled by Lennard-Jones potential, results subdiffusive dynamics at the intermediate time for the passive or weakly active tracer, but superdiffusive for moderately to the strongly active tracer. However, in all the cases, the long-time dynamics is purely diffusive. The intermediate subdiffusive behavior is a result of local trapping events of the tracer, which is pronounced for the passive tracer or if the tracer is only weakly active. For higher self-propulsion, this intermediate time subdiffusive behavior changes to superdiffusive.
 Another aspect of the tracer dynamics is its deviation from Gaussianity. The deviation is only for the intermediate time and is reflected in the value of the non-Gaussian parameter, which is zero for the Gaussian case but non-zero if it is not Gaussian.
 The intermediate time non-Gaussian behavior of the tracer dynamics is a combined result of excluded volume interaction with the chain, trapping (when sticky) offered by the chain monomers. In other words, it is caused by the heterogeneity of the medium. However, if one waits sufficiently long, the Gaussianity is recovered. For the sticky, passive tracer, the intermediate time non-Gaussianity is more pronounced for moderate stickiness, as then the tracer feels the free space as well as the polymer chain. For higher stickiness, the tracer mostly remain close to the chain and therefore feels a restricted but less heterogeneous environment. {The case of self-propelled tracer is even more interesting. For moderate self-propulsion $\text{NGP}$ is increases to more positive values but for higher F it becomes negative as the persistent motion dominates over the random diffusion. At lower activity, tracer stays in binding zones and at higher F it escapes easily. We believe this observation is very important and useful while one engineers self-propelled vehicles for drug delivery. Our simulations indicates that though higher self-propulsion makes the tracer to reach its target faster, it also accelerates escape dynamics of the tracer from the trap. Thus, ideally there has to be an optimum self-propulsion velocity which will make the dynamics faster than the passive case but also allows sufficient time for the tracer to remain bounded to the target.}  For the sticky tracer, trapping events are inevitable, and it is long-lived when the stickiness is high. This can be seen from the plots of the trapping time distributions, which are broader for higher stickiness ($\epsilon$). On the other hand, increasing self-propulsion helps the tracer to escape from these traps and then the trapping time distribution becomes narrower with increasing self-propulsion force. The size and the dynamics of the tracer particle also have profound consequences on the conformation of the trapping zone of the polymer chain, which is evident from the radius of gyration calculation. Our analyses show that for the passive as well as for the self-propelled tracer particle, $R_{g}$  of the trapping zone of the chain decreases, pointing to the collapse of the zone with the increase of the size of the tracer particle. For the self-propelled tracer, $R_g$ of the trapping zone is higher in comparison to a passive one with the same size, owing to the faster dynamics of the self-propelled tracer.\\

\noindent In a nutshell, our present work focuses on the complex dynamics of either self-propelled or a passive tracer particle in a polymeric medium. {There is a class of phenomena where events involving self-propelled tracer particles diffusing and binding to macromolecules play central role. {Examples include binding unbinding events of ligands with polymer as binding of transcription factor to DNA, where a tracer particle (transcription factor) binds to a polymer (DNA) \cite{garcia2021power}, functions of enzymes where substrate binds to the specific zone (active site) of enzyme and many biological functions, and engineered self-driven micron or nano-sized devices for targeted drug delivery.\cite{patra2013intelligent}} In general, targeted delivery of the drug facilitates by presence of activity, which helps the drug to navigate the path and faster delivery.\cite{chen2019magnetic} In our current work, the polymer chain has only one binding zone for the tracer, but the inclusion of multiple binding zones of different trapping strengths seems straightforward.} We believe that our current study will be useful to scientists and engineers to design efficient artificial self-driven transport machines that can function in crowded environments such as biological cells. 

\section*{Conflicts of interest}
There are no conflicts to declare.

\section*{Acknowledgements}
\noindent R. S. Y. thanks IIT Bombay for the fellowship. R. C. acknowledges SERB, India, Project No. MTR/2020/000230 under the MATRICS scheme and IRCC-IIT Bombay (Project No. RD/0518-IRCCAW0-001) for funding. R. S. Y. and C. D. acknowledge Dr. Koushik Goswami, Praveen Kumar, and Rajiblochan Sahoo and special thanks to Ligesh Theeyancheri for helpful discussions. Authors thank Sanaa Sharma and Pooja Nanavare for reading the manuscript. We acknowledge the SpaceTime-2 supercomputing facility at IIT Bombay for the computing time.

\onecolumngrid \section*{\LARGE{Supplementary Material}}

\renewcommand{\arraystretch}{2.0}  
\begin{table}[h]
\small
  \caption{Model Parameters }
  \label{tb2:example2}
  \begin{tabular*}{0.48\textwidth}{@{\extracolsep{\fill}}|c|c|}
    \hline
    \textbf{Parameter} & \textbf{Value}\\
    \hline
    $\sigma\textsubscript{tracer}$ & 1-4  \\
    $\sigma_{\text{beads}}$ & 1  \\
    $\frac{m}{\zeta}$ & $10^{-3}$  \\
    $k_{B}T$& 1 \\
    $\Delta t$ & $5\times 10^{-4}$ \\
     F & 0, 8, 12, 16, 20, 40 \\
      $\epsilon$ & 1, 2, 3, 4, 6 \\
    \hline
  \end{tabular*}
\end{table}

\clearpage
\renewcommand{\thefigure}{S\arabic{figure}}
\setcounter{figure}{0}
\begin{figure*} [h!]
    \includegraphics[width=0.95\linewidth]{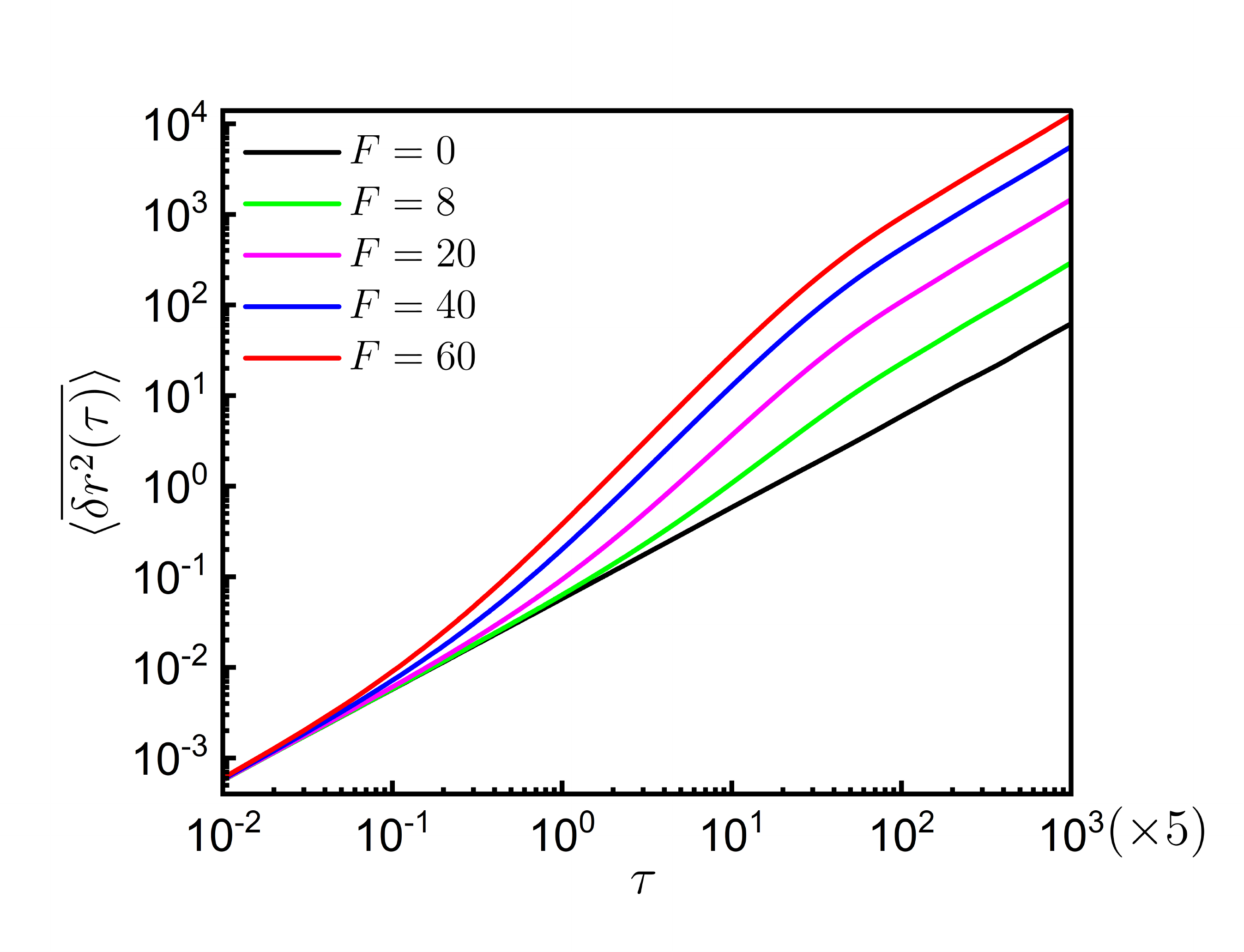}
    \caption{ Log-log plot of $\left<\overline{\Delta r^{2}(\tau)}\right>$ vs $\tau$ for free particle with different F.}
    \label{fig:S1_free_particle_msd_dist}
\end{figure*}

\begin{figure*} [h!]
    \includegraphics[width=0.95\linewidth]{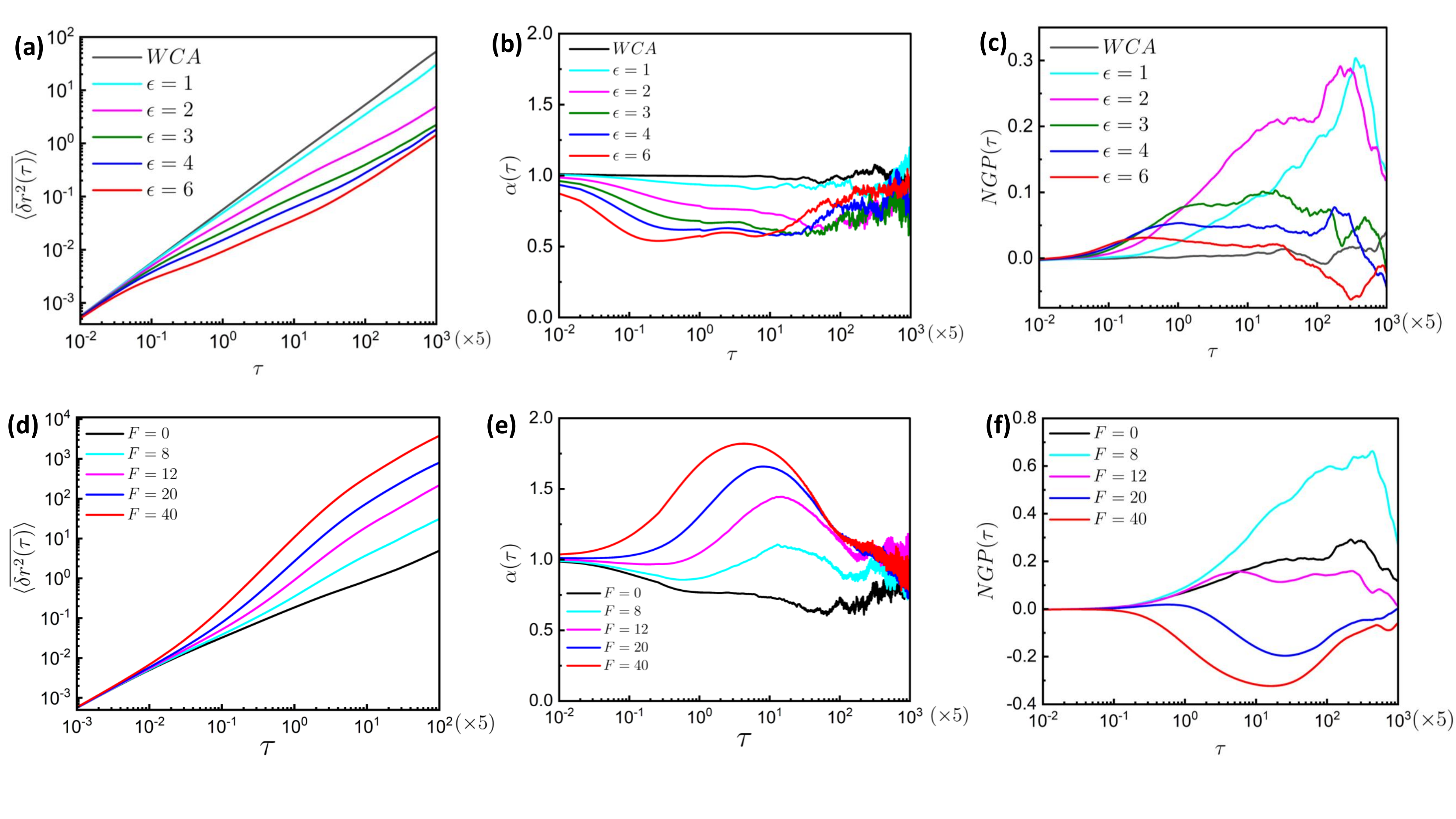}
    \caption{Log-log plot of $\langle\overline{\delta r^2(\tau)}\rangle$ for (a) different $\epsilon$ (F = 0), (d) different F $(\epsilon = 2)$, log-linear plot of $\alpha(\tau)$ for (b) different $\epsilon$ at F = 0, (e) different F at $\epsilon = 2$, NGP($\tau$) for(c) different $\epsilon$ at F = 0, and (f) different F for $\epsilon = 2$. ( N = 200).} 
    \label{fig:S1_free_particle_msd_dist}
\end{figure*}

\begin{figure*} [h!]
    \includegraphics[width=0.95\linewidth]{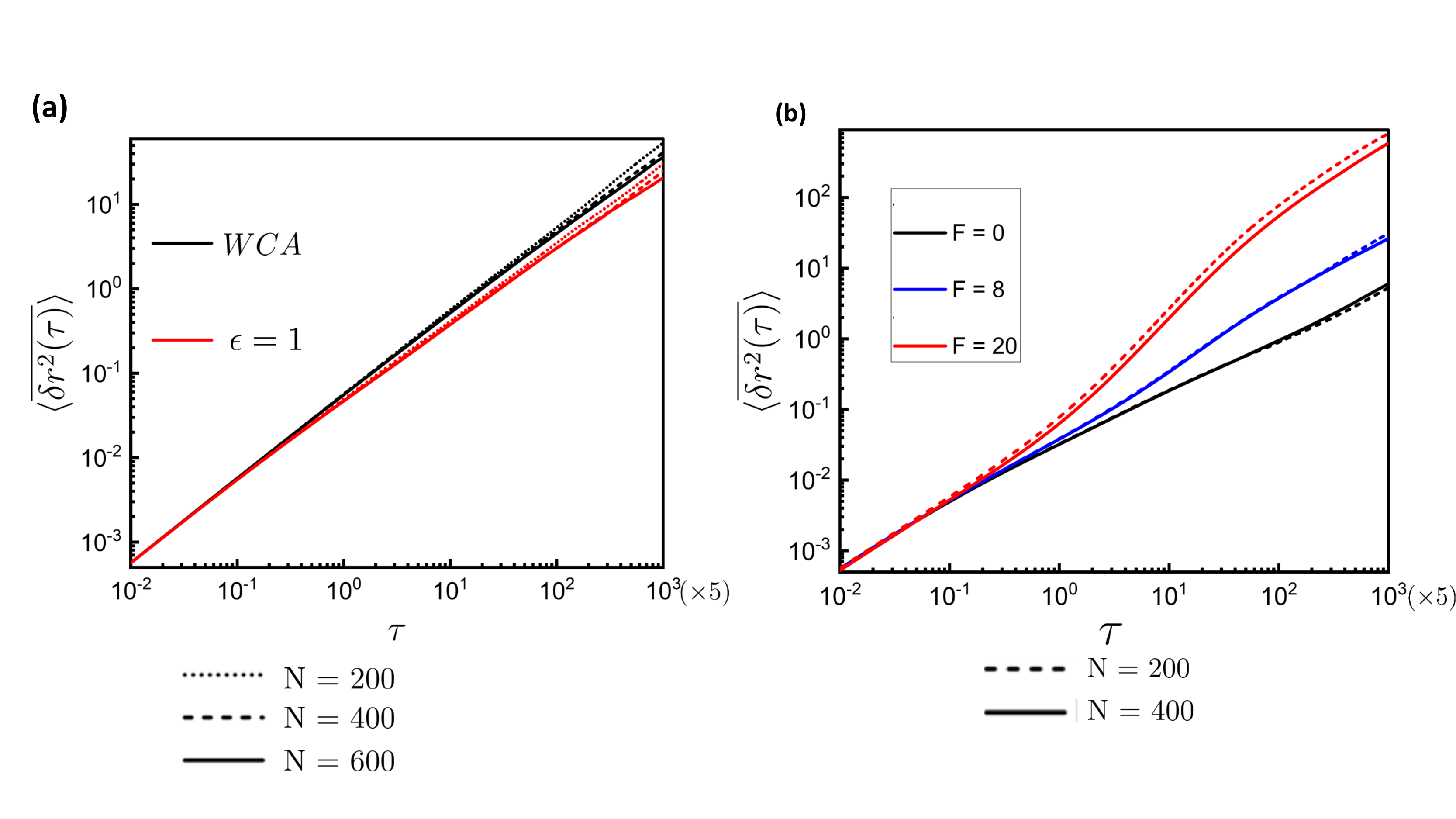}
    \caption{Log-log plot of $\langle\overline{\delta r^2(\tau)}\rangle$ for (a) different $\epsilon$ (F = 0), (b) different F $(\epsilon = 2).$}
    \label{fig:S1_free_particle_msd_dist}
\end{figure*}

\begin{figure*} 
    \includegraphics[width=1.0\linewidth]{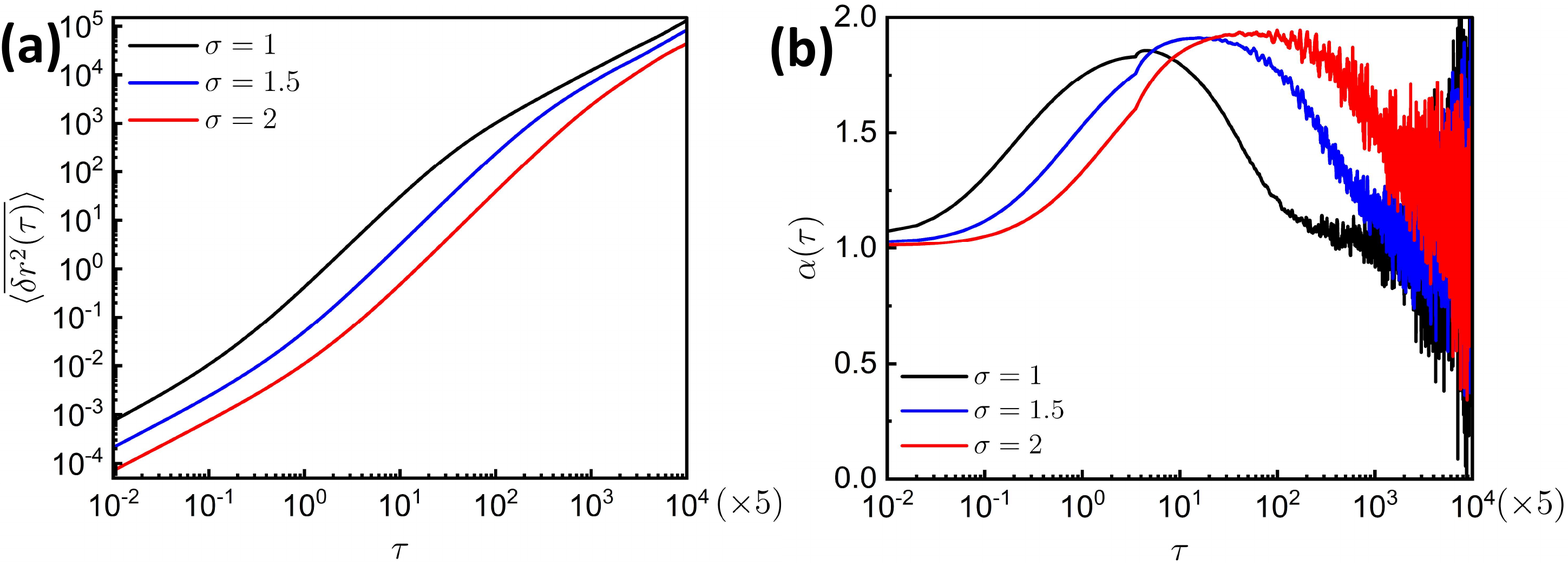}
    \caption{(a) Log-log plot of $\left<\overline{\delta r^{2}(\tau)}\right>$ and (b) log-linear plot of $\alpha(\tau)$ vs $\tau$ for different size of the tracer with constant $F=40$, $\epsilon=2$.}
    \label{fig:S2_varrying_sigma_act}
\end{figure*}

\begin{figure*}
    \includegraphics[width=0.6\linewidth]{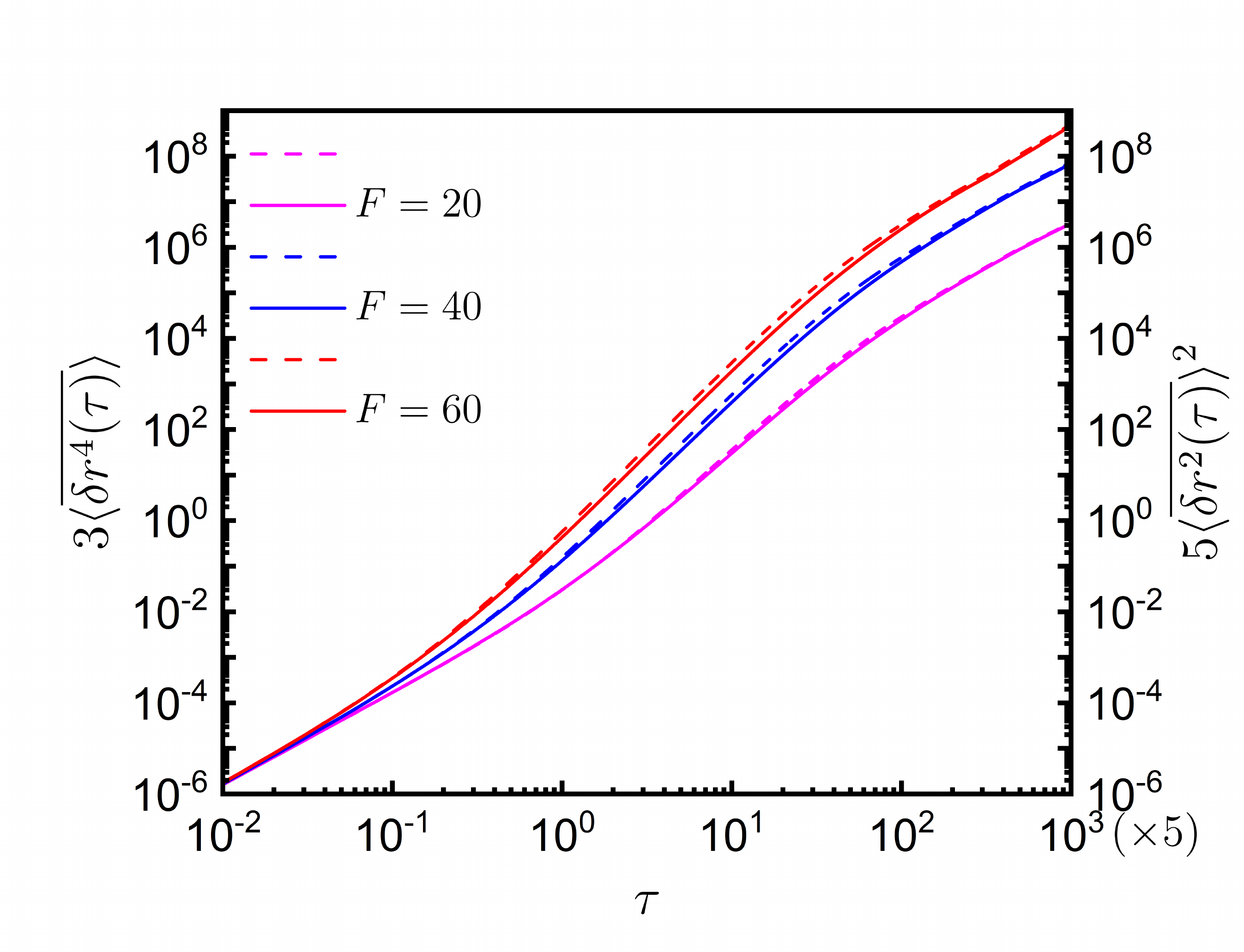}
    \caption{Log-log plot for fourth moment of displacement $3\langle\overline{\delta r^4(\tau)}\rangle$ (solid line) and square of mean square displacement $5\langle\overline{\delta r^2 (\tau)}\rangle^2$ (dotted line) vs $\tau$ for different F and $\epsilon=2.$ }
    \label{fig:S3_pdf_varry_tau}
\end{figure*}

\begin{figure*} [h!]
    \includegraphics[width=0.95\linewidth]{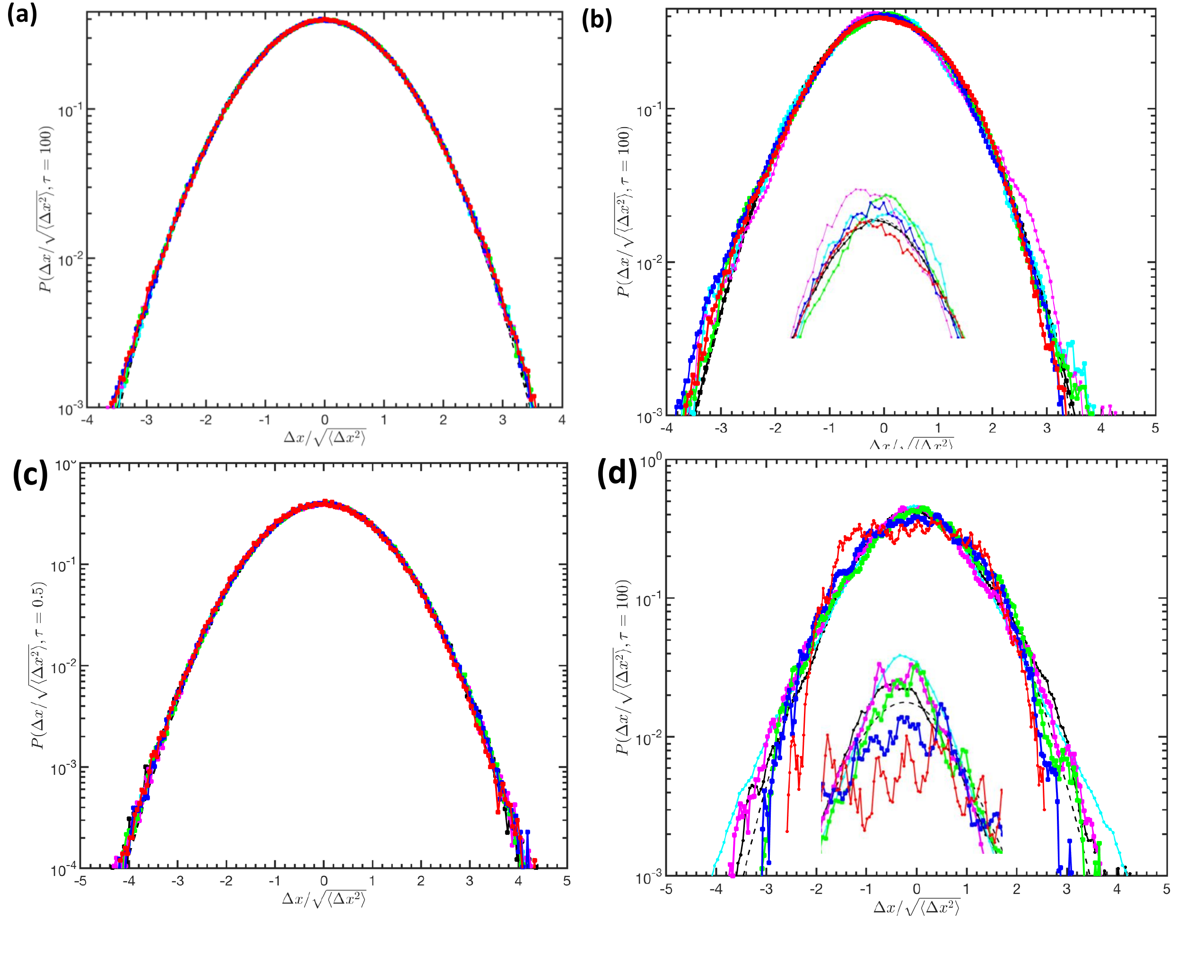}
    \caption{Log linear plot of $P(\Delta X; \tau = 0.5)$ for (a) different $\epsilon$ at F = 0, (b) different F at $\epsilon = 2$, and $P(\Delta X; \tau = 100)$ for (c) different $\epsilon$ at F = 0, (d) different F at $\epsilon = 2$. Inset one is the enlarge peak of the distribution.}
    \label{fig:S1_free_particle_msd_dist}
\end{figure*}

\begin{figure*} [h!]
    \includegraphics[width=0.95\linewidth]{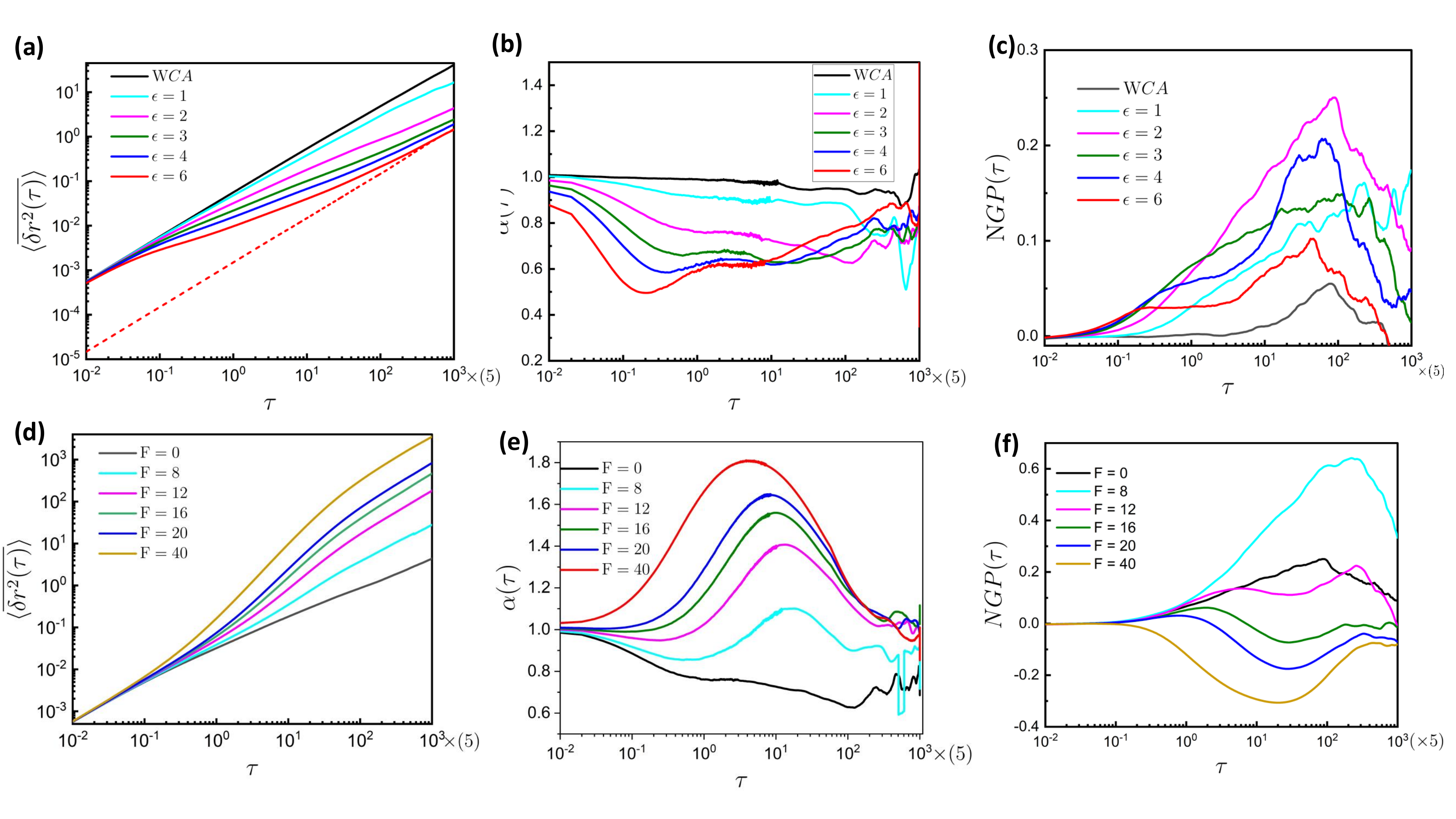}
    \caption{Log-log plot of $\langle\overline{\delta r^2(\tau)}\rangle$ for (a) different $\epsilon$ (F = 0), (d) different F $(\epsilon = 2).$, log-linear plot of $\alpha(\tau)$ for (b) different $\epsilon$ at F = 0, (e) different F at $\epsilon = 2$, NGP$\tau$ for(c) different $\epsilon$ at F = 0, and (f) different F for $\epsilon = 2$ (N = 400, polymer with specific binding zone).}
    \label{fig:S1_free_particle_msd_dist}
\end{figure*}

\begin{figure*} [h!]
    \includegraphics[width=0.95\linewidth]{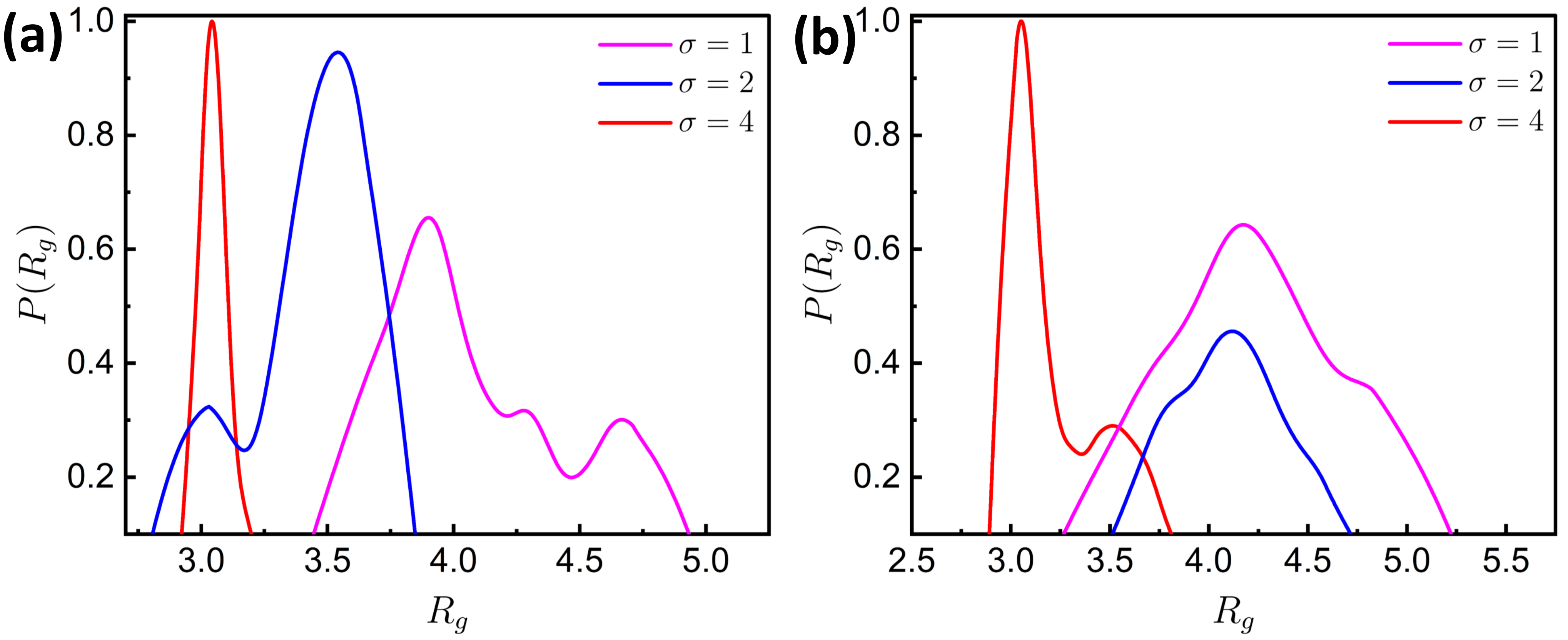}
    \caption{Probability distribution of $R\textsubscript{g}$ of polymer trapping zone for (a) passive tracer particle $F=0$, $\epsilon=2$ and (b) self-propelled tracer particle $F=20$, $\epsilon=2.$ (N = 200, polymer with specific binding zone.)} 
    \label{fig:S1_free_particle_msd_dist}
    
\end{figure*}

\begin{figure*} [h!]
    \includegraphics[width=0.95\linewidth]{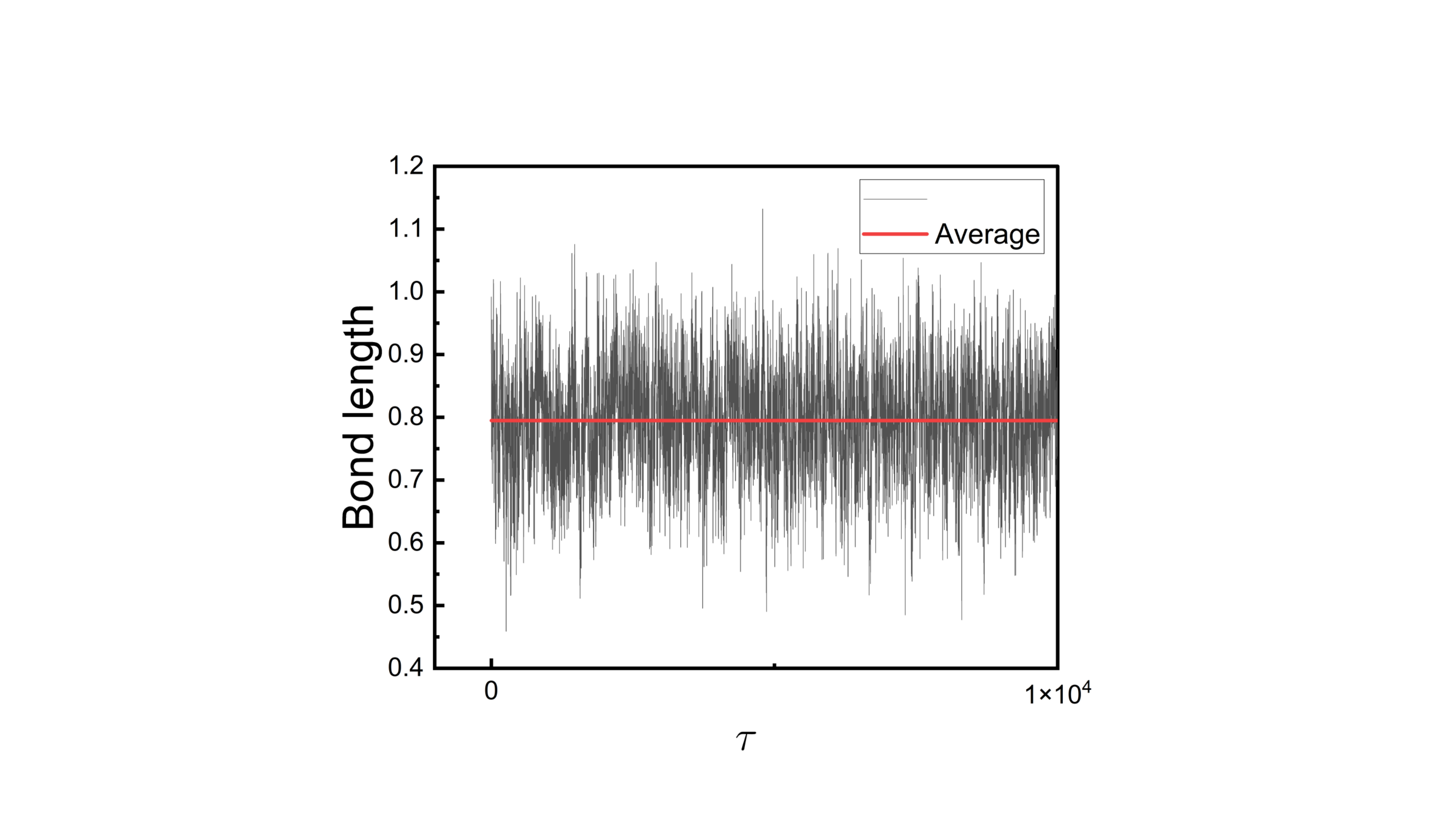}
    \caption{plots of average bond fluctuation with time}
    \label{fig:S1_free_particle_msd_dist}
\end{figure*}

\clearpage
\twocolumngrid \subsection{Movie Description}

\noindent \textbf{Movie description} \\
\noindent Movie\_S1 \\
\noindent Molecular dynamics simulation of the passive tracer particle (red) $F=0$ in the polymeric medium consist of all sticky beads (green in color). It is clear from the movie that the passive tracer particle feel confinement by the polymer chain  ($N = 400$) and the dynamics is slow.\\

\noindent Movie\_S2 \\
\noindent Molecular dynamics simulation of the self-propelled particle (red) $F=8$ in the polymeric ($N = 400$) medium consist of all sticky beads (green in color). We observe that for the moderate self-propelled force propulsion enhance the dynamics of tracer particle but not necessary to overcome the stickiness and crowding effect. \\

\noindent Movie\_S3 \\
\noindent Molecular dynamics simulation of the self-propelled particle (red) $F=20$ in the polymeric ($N = 400$) medium consist of all sticky beads (green in color) with all sticky beads. Here, we observe that for high self propulsion tracer particle can escape from the crowing and stickiness effect. \\

\noindent Movie\_S4 \\
\noindent Molecular dynamics simulation of the passive tracer particle (red) with $\sigma_{tracer}=1$, $(F=0)$ in the polymeric ($N = 400$) medium consist of 200 sticky beads at center (green) and others are repulsive (blue). Here, the tracer particle spends maximum time in the trapping zone.\\

\noindent Movie\_S5 \\
\noindent Molecular dynamics simulation of the self-propelled tracer particle (red) with $\sigma_{tracer}=1$, $(F=8)$ in the polymeric ($N = 400$) medium consist of 200 sticky beads at centre (green) and others are repulsive (blue). Here the particle escapes from the trapping zones due to self-propulsion.\\

\noindent Movie\_S6 \\
\noindent Molecular dynamics simulation of the self-propelled tracer particle (red) with $\sigma_{tracer}=4$, $(F=20)$ in the polymeric ($N = 400$) medium consist of 200 sticky beads at centre (green) and others are repulsive (blue). The trapping zones of polymer is getting collapse due to strong sticky interactions of tracer particle. \\

\noindent Movie\_S7 \\
\noindent Molecular dynamics simulation of the passive tracer particle (red) with $\sigma_{tracer}=4$, $(F=0)$ in the polymeric ($N = 200$) medium consist of 100 sticky beads at centre (green) and others are repulsive (blue). The trapping zones of polymer is getting collapse  and fluctuate due to  larger size of tracer particle and activity of the tracer particle. \\

\noindent Movie\_S8 \\
\noindent Molecular dynamics simulation of the self-propelled tracer particle (red) with $\sigma_{tracer}=4$, $(F=20)$ in the polymeric ($N = 200$) medium consist of 100 sticky beads at centre (green) and others are repulsive (blue). The trapping zones of polymer shows less collapse and fluctuate due to  larger size of tracer particle and activity of the tracer particle.

\clearpage
\providecommand*{\mcitethebibliography}{\thebibliography}
\csname @ifundefined\endcsname{endmcitethebibliography}
{\let\endmcitethebibliography\endthebibliography}{}

\end{document}